\documentclass{aa}

\usepackage[colorlinks=true,allcolors=blue]{hyperref}
\usepackage[varg]{txfonts}
\usepackage{amsmath}
\usepackage{mathtools}
\usepackage{graphicx}
\usepackage{xcolor}
\usepackage{array}
\usepackage{gensymb}
\usepackage{tablefootnote}
\usepackage{lipsum}
\usepackage{ulem}
\usepackage{comment}
\usepackage{tabularx}
\usepackage{orcidlink}
\usepackage{diagbox}
\usepackage{lineno}
\nolinenumbers\renewcommand{\linelabel}[1]{}

\begin{document}

\definecolor{forestgreen}{rgb}{0.13, 0.55, 0.13}

\newcommand{\todo}[1]{\textcolor{cyan}{\bf Todo: #1}}
\newcommand{\com}[1]{\textcolor{forestgreen}{#1} }
\newcommand{\rev}[1]{#1}
\newcommand{\revnew}[1]{#1}

\title{Asymmetric signatures of warps in edge-on disks
}
\author{Carolin N. Kimmig\inst{1,2}\orcidlink{0000-0001-9071-1508} \and
        Marion Villenave\inst{2, 3}\orcidlink{0000-0002-8962-448X}}

\institute{
	 Zentrum f\"ur Astronomie, Heidelberg University, Albert-Ueberle-Str.~2, 69120 Heidelberg, Germany
  \and Dipartimento di Fisica, Università degli Studi di Milano, Via Giovanni Celoria 16, 20133 Milano, Italy
  \and Univ. Grenoble Alpes, CNRS, IPAG, 38000 Grenoble, France
}

\date{Received xxx / Accepted xxx} 

\abstract {
Protoplanetary disks observed edge-on commonly show asymmetries that can be caused by shadows cast from a misaligned inner region or warp.
With the growing amount of these observations, methods to constrain warp parameters are urgently needed.
}
{In this work, we investigate observational signatures of warps in edge-on disks with the aim to find limits on the observability of warps.
}
{We perform radiative transfer simulations in scattered light (near-/mid-infrared) of edge-on disks containing warps of different amplitudes and varying sizes of the misaligned inner region. We analyze the effect of these parameters on observations at $\lambda~=~0.8\,\mathrm{\mu m}$, but also compare models for specific parameters at wavelengths up to $\lambda~=~21\,\mathrm{\mu m}$.
In all models, we quantify the asymmetry by fitting the two nebulae corresponding to the disk surfaces individually.
}
{We find that under optimal conditions, an asymmetry due to slight warps can be visible.
The visibility depends on the viewing angle, as warped disks are not axisymmetric.
For optimal azimuthal orientation, misalignments between inner and outer disk regimes as low as $2^\circ$ can lead to significant asymmetries.
On the other hand, we find that larger \revnew{misalignments} of about $10^\circ$ are needed in order to observe a change in the brightest nebula as a function of wavelength, which is observed in a few protoplanetary disks.
In order to link our results to observations, we investigate potential misalignments between the inner and outer disk of edge-on systems showing asymmetries. We compare the jet geometry, which presumably traces the very inner regions of the disk, to the outer disk orientation of four systems and find that a misalignment of a few degrees could be consistent with the data, such that warped disks could explain the apparent lateral asymmetries.
}
{There are many factors that can influence the appearance of the shadows and asymmetries in warped disks. Although it is still challenging to infer exact warp parameters from observations, this work presents a promising step toward better constraints.
}

\maketitle 

\section{Introduction}

Understanding the physical structure of protoplanetary disks is of critical importance to understand planet formation. 
Historically, protoplanetary disks have been assumed to be axi-symmetric with all material distributed along the same plane. 
However, a growing number of observations is revealing the presence of asymmetric features, such as spirals, shadows, and other lateral asymmetries~\citep[e.g., reviews by][]{Benisty_2023, Bae_2023}. Some of these asymmetries can be indications of a misaligned inner region or warp casting a shadow onto the outer part of the disk. Misalignments or warps might in fact be fairly common in protoplanetary disks.
A local misalignment (i.e., broken disk) or warp (i.e., smooth misalignment) will locally perturb the planet formation process and can be a cause for the diversity of planets.

Warps can be formed in many different scenarios.
One scenario is a bound companion object, either a planet \citep{Facchini2014} or a binary companion which is inclined with respect to the disk plane. This could either be an outer companion \citep{ZanazziLai2018a}, a system with a circumbinary disk \citep{Facchini2013, Lodato2013, ZanazziLai2018b, Deng2022, Rabago2023, Young2023}, or any combination of both.
This companion object exerts a gravitational torque on the disk. This torque causes concentric rings within the disk to precess. Because the torque highly depends on the distance to the gravitational perturber, the concentric rings precess on different timescales. This produces the warp shape.
The same principle applies to a flyby of a massive object inclined with respect to the disk \citep[][Kimmig et al., in prep.]{Picogna2014, Dai2015, Xiang-Gruess2016, Cuello2019, Cuello2023}. Such fly-bys can be common, especially in the early phases of star and disk formation \citep{Pfalzner2021}.
A further formation scenario is late infall of material, which can add angular momentum in a different direction and thus distort the disk \citep{Dullemond2019, Ginski2021, Kuffmeier2021, Krieger2024}.
Finally, a change in the rotation axis during the formation process due to a magnetic field is also suggested to cause misalignments \citep[AA Tau;][]{Bovier1999}.

The diversity of formation scenarios already hints at the variety of appearances of misalignments in observations.
Additionally, warps are highly dynamic \citep{Papaloizou1983, Papaloizou1995}.
{Once the warp is formed through one of the formation scenarios, internal torques come into play. These internal torques arise} due to the misalignment of neighboring orbits, which leads to pressure gradients in the disk.
The pressure gradients then cause resonant motions of the gas, called sloshing motions \citep{Dullemond2022}, and vertical breathing modes \citep{Held2024}.
These motions exert a torque on the different orbital planes within the disk and hence change the orientation of the orbits.
{Thus, the net effect of the internal torque and pressure gradients is the evolution of the warp shape.}

In typical conditions of a protoplanetary disk (especially low viscosity), the warp travels as a wave through the disk, often referred to as the wave-like regime \citep{LubowOgilvie2000, Gammie2000, OgilvieLatter2013a, OgilvieLatter2013b}.
Its wave speed is related to the sound speed in the disk \citep{NixonKing2016}.
The wave is dampened over time, mainly by viscosity, on a damping time scale that can be estimated by $t_\mathrm{damp} \approx 1/(\alpha \Omega_\mathrm{K})$, with $\Omega_\mathrm{K}$ as Kepler frequency evaluated at the outer regime of the disk \citep{LubowOgilvie2000}. {Here, $\alpha$ is the Shakura-Sunyaev viscosity parameter \citep{ShakuraSunyaev1973}.} This time scale typically ranges from $10^3$ to $10^5\,\mathrm{years}$ \citep{Foucart2014, Poon2021, Rowther2022, Kimmig2024}.
{During their evolution, warps may also break into separate disks under certain conditions \citep{Nixon2013, Dogan2023, Young2023, Rabago2024}}.

Because of their dynamic evolution, warps can take on various shapes leading to a plethora of disk misalignment observations.
Indeed, asymmetric observations linked to shadows due to warps are recurrent in scattered light imaging of low to mid-inclination protoplanetary disks~\citep[e.g.,][]{Garufi2018, Garufi2022, Benisty_2023}. 
Low-inclination systems can show narrow dark lanes \citep{ Benisty2017, Pinilla2018, Keppler2020}, broad dark regions \citep{Garufi2016, Debes2017, Muro-Arena2020, Kraus2020, Hashimoto2024}, or a combination of both \citep{Stolker2016, Benisty2018}.
Furthermore, the sensitivity and spectral resolution of ALMA are now allowing to reveal a larger number of warped disks, for example via gas kinematic observations~\citep[e.g.,][]{Casassus2015, Loomis_2017, vanderPlas_2017, Perez_2018, Sakai2019, Yamato2023}.

In parallel, more and more studies also aim to directly compare the orientation of the inner and the outer regions of protoplanetary disks. \cite{Francis_2020} compared the inclination and position angle of the inner disk ($\lesssim 20$au) and main ring of 14 transition disks observed at high angular resolution with ALMA, finding that 8 of them are significantly misaligned. \cite{Ansdell2020} instead analyzed a sample of 24 dipper stars, for which the inner disk is believed to have an inclination close to edge-on, and found that the outer disks follow a random distribution of orientations, indicative of significant misalignment for these systems. \cite{Bohn2022} compared the inclination of the outer disk, based on ALMA observations, with that of the inner disk, obtained using VLT/Gravity for a subsample of 20 transition disks. They identify 6/20 disks with clear evidence of misalignments, all of which showing shadows on their outer disks at scattered light wavelengths, compatible with the derived inner disk geometry.
\cite{Kluska2020} use reconstructed images of near-infrared interferometry observations to evaluate the symmetry of the innermost region of 15 protoplanetary disks around Herbig AeBe stars. They find evidence of non-axisymmetric emission that can be linked to warps in 5/15 disks.

A key to interpreting the plethora of misalignment observations are models for identifying signatures of warped disks.
Such models are able to provide kinematic insight \citep{Casassus2015, Young2022, Zuleta2024} or chemical signatures \citep{Young2021} in disks containing warps.
Shadows can be investigated in synthetic observations in scattered light \citep{Nealon2019, Juhasz2017}.
\cite{Fachhini2018} investigated shadow signatures of broken disks in intermediate inclination disks and were able to reproduce both dark lanes for strong misalignments, and broad regions for low to moderate misalignments.
The location of observed shadows can be used to constrain the geometry of the inner, misaligned disk \citep{Min2017}.
The shadows can be variable, e.g. for precessing broken disks \citep{Nealon2020}.

Asymmetries can also occur in disks observed edge-on \citep{Villenave2024}.
Edge-on disks typically appear in scattered light observations as two nebulae separated by a dark line~\citep[e.g.,][]{Burrows1996, Stapelfeldt1998}.
Those two nebulae correspond to the surface regions of the disk, while the dark line occurs because the material close to the midplane obscures the light from the host star.
Thus, the characteristics of the dark line highly depend on properties and vertical distribution of the dust contained in the disk \citep{Watson_2007}.
On the other hand, the surfaces of the disk are less dense and therefore multiple scattering events allow photons to escape the disk and reach the observer, thus leading to the two bright nebulae.
In a perfectly planar, unwarped disk, viewed perfectly edge-on, the nebulae would appear entirely symmetric.
Even though observing such a disk with a slight inclination variation would lead to one of the nebulae being brighter due to forward scattering, the extent of both nebulae along their major axis would still be symmetric.
However, observations show that a large fraction of edge-on protoplanetary disks at scattered light wavelengths 
display a shift or lateral asymmetry of their nebulae~(15/20 disks analyzed by \citealt[][]{Villenave2024}, see also \citealt[][]{Berghea_2024}).
A misalignment of the inner disk region is proposed to explain this asymmetry, as reproduced in some models \citep{Fachhini2018, Nealon2019, Villenave2024}.

In this work, we specifically target the scattered light appearance of edge-on and almost edge-on disks containing a warp.
We aim to investigate how the asymmetry of the two nebulae changes depending on the warp morphology and to find limits of observability.
In addition, we also evaluate the warp geometry that can lead to a switch in the brightest nebula with wavelength, which has been observed in a few systems \citep{Grosso2003, Villenave2024}.
We present our models in Section~\ref{sec:nummodels} and the results in Section~\ref{sec:results}.
In Section~\ref{sec:jets}, we link our results to observations by evaluating the misalignment between the outer regions of edge-on disks and jets presumably linked to the inner region.
We conclude our work in Section~\ref{sec:conclusion}.

\section{Methods} \label{sec:nummodels}

To study the appearance of warped edge-on protoplanetary disks, we perform radiative transfer simulations using RADMC-3D \citep{radmc3d}.
We create most models in a wavelength of $\lambda~=~0.8\,\mathrm{\mu m}$, but extend it to $\lambda~=~2\,\mathrm{\mu m}$, $12\,\mathrm{\mu m}$, and $21\,\mathrm{\mu m}$ for some parameter sets.
{Images in this near- to mid-infrared regime are often called scattered light images, as scattered light usually dominates the appearance of disks in these wavelengths.
In the longer, mid-infrared regime wavelengths thermal emission can contribute to the radiation from disks, and we include the thermal emission in all of our models. However, even with the thermal emission, the scattered light scattered light dominates the outcome of the images.}
The images are scaled by $2.28\,\mathrm{au}$ per pixel, and for the evaluation using spine fitting, we assume a distance of $140\,\mathrm{pc}$.

\subsection{Model set-up}

We set up an initial gas density structure, where the surface density profile follows a power law with an exponential cut-off both at the inner and outer disk edge, according to
\begin{equation}
\Sigma(r) = \Sigma_0 \left( \frac{r}{R_0} \right)^{-p} \left(\frac{1}{1 + \exp(\frac{r - r_\mathrm{out}}{0.05\ r_\mathrm{out}})} \right) \left( \frac{1}{1 + \exp(\frac{r_\mathrm{in} - r}{0.05\ r_\mathrm{in}})} \right),
\end{equation}
with $r_\mathrm{in}$ and $r_\mathrm{out}$ as inner and outer disk edge, respectively, and $\Sigma_0$ as surface density at the reference radius $R_0$, which we set to $R_0~=~5.2\,\mathrm{au}$.
According to typical temperature structures in protoplanetary disks, the vertical extent can be described by
\begin{equation}
h(r) = \frac{H_\mathrm{p}}{r} =  h_0 \left(\frac{r}{R_0} \right)^{i_\mathrm{fl}},
\end{equation}
where $h_0$ is the aspect ratio at the reference radius $R_0$ and $i_\mathrm{fl}$ is the flaring index.
To parametrize the warp, we implement a tilt profile inspired by \cite{Martin2019} of
\begin{equation} \label{eq:warp-parametrization}
i(r) = i_\mathrm{max} \left[\frac{1}{2} \tanh \left( \frac{r - r_\mathrm{warp}}{r_\mathrm{width}} \right) - \frac{1}{2} \right],
\end{equation}
with $i_\mathrm{max}$ as maximum warp tilt, $r_\mathrm{warp}$ the warp location, and $r_\mathrm{width}$ is the width of the warp transition.
This setup is identical to \cite{Kimmig2024}, with adjusted parameters.

In all our models, the disk ranges from $r_\mathrm{in}=2\,\mathrm{au}$ to $r_\mathrm{out}~=~200\,\mathrm{au}$, while the grid extends from $0.5\,\mathrm{au}$ to $260\,\mathrm{au}$ in the radial direction.
We use a full spherical grid, i.e., the vertical extent is set from $0$ to $\pi$.
For the resolution, we chose $250$ grid cells in the radial direction, $100$ cells in the azimuthal direction and $250$ cells in the vertical direction.
We set the aspect ratio to $h_0 = 0.1$ and use a flaring disk geometry with $i_\mathrm{fl}=0.1$. Similar parameters have been obtained from the modeling of scattered light images of several disks~(e.g., \citealt{Muro-Arena_2018, Sturm2023}, see also \citealt[][]{Angelo_2023}).

We choose for the star's properties a mass of $M_* = 2.5\,M_\odot$, a radius of $R_* = 1.8\,R_\odot$, and an effective temperature of ${T = 10500\,K}$.
For the gas distribution, we set a slope of $p=1$ and $\Sigma_0~=~34.3\,\mathrm{g/cm^2}$ at the reference radius of $R=5.2\,\mathrm{au}$, resulting in a gas disk mass of $M_\mathrm{gas}=0.01\,M_*$.

\begin{figure}[ht!]
\centerline{\includegraphics[width=0.5\textwidth]{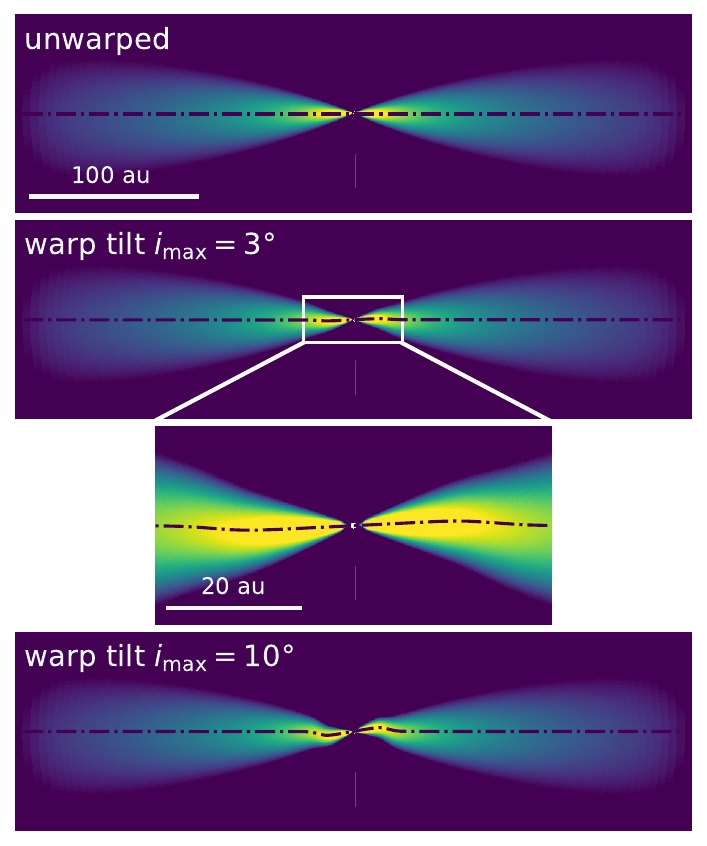}}
     \caption{\label{fig:fargo-wampl} Vertical cross-section of the gas density set-up for an unwarped disk, a warp of $3\degree$, and a warp of $10\degree$. The dash-dotted line indicates the disk midplane.}
\end{figure}

In order to investigate the effect of differently warped disks, we vary the warp tilt $i_\mathrm{max}$ from $0\degree$ to $10\degree$ in steps of $1\degree$.
For the location of the warp, we set a fiducial value of $r_\mathrm{warp}~=~20\,\mathrm{au}$, but vary the location of the warp for a warp tilt of $3\degree$ to $r_\mathrm{warp}~\in~\{5\,\mathrm{au},50\,\mathrm{au},100\,\mathrm{au}\}$.
To get a smooth warp transition at all locations, we scale the warp width with the location of the warp using $r_\mathrm{width}~=~0.25\,r_\mathrm{warp}$.
{We can measure the warp strength in each case by evaluating the maximal displacement of the disk midplane in terms of the pressure scale height at the location of the warp. 
Equation~\ref{eq:warp-parametrization} shows that the tilt angle between the disk midplane at the warp location $r_\mathrm{warp}$ with respect to the inner (as well as outer) disk midplane is $i_\mathrm{max}/2$. We can then calculate the displacement $w$ from the height of the midplane $z_\mathrm{mid}(r_\mathrm{warp})$ with respect to the midplane of the inner (or outer) disk}
{\begin{equation}\label{eq:warpstrength}
    w = \frac{z_\mathrm{mid}(r_\mathrm{warp})}{H_\mathrm{p}(r_\mathrm{warp})} = \frac{r_\mathrm{warp} \tan \left(\frac{i_\mathrm{max}}{2} \right)}{H_\mathrm{p}(r_\mathrm{warp})}.
\end{equation}}
{The numbers of $w$ for the cases in our investigation are given in Table~\ref{tab:warpstrength} and } Figure~\ref{fig:fargo-wampl} shows vertical cross-sections through the gas density for three example cases with different warp {tilts (for $r_\mathrm{warp}=20\,\mathrm{au}$).} 
\begin{table}[ht!]
    \centering
    \caption{{Overview over the warp strength $w$ in our disk set-ups}}
    \footnotesize
    {
    \begin{tabular}{p{0.75cm}|p{0.34cm}p{0.34cm}p{0.34cm}p{0.34cm}p{0.34cm}p{0.34cm}p{0.34cm}p{0.34cm}p{0.34cm}p{0.34cm}} \label{tab:warpstrength}
     \diagbox[width=1.2cm]{$r_\mathrm{warp}$}{$i_\mathrm{max}$} & $1^\circ$ & $2^\circ$ & $3^\circ$ & $4^\circ$ & $5^\circ$ & $6^\circ$ & $7^\circ$ & $8^\circ$ & $9^\circ$ & $10^\circ$  \\ \hline
     $5\,\mathrm{au}$ & & & 0.26 & & & & & & & \\
    $20\,\mathrm{au}$ & 0.08 & 0.15 & 0.23 & 0.31 & 0.38 & 0.46 & 0.53 & 0.61 & 0.69 & 0.76 \\
    $50\,\mathrm{au}$ & & & 0.21 & & & & & & & \\
    $100\,\mathrm{au}$ & & & 0.19 & & & & & & & \\
    \end{tabular}
    }
\end{table}

\subsection{Radiative transfer}

For the radiative transfer simulations, we need to include dust in our models.
We assume small dust particles, which are perfectly coupled to the gas and well-mixed, which means that the structure of the dust perfectly follows the structure of the gas disk.
Then, we chose a dust-to-gas ratio of $10^{-3}$, leading to a total dust mass of $M_\mathrm{dust} = 10^{-5} M_*$.
We note that this dust-to-gas ratio is slightly lower than typically assumed, because small dust particles likely only make up a fraction of the total dust mass in protoplanetary disks.
Despite recent studies on the behavior of larger dust particles in warped disks \citep{Cuello2019, Aly2021, Aly2024}, many aspects remain unexplored, which is why we decided to only consider {a single} grain size of $a=10\,\mathrm{\mu m}$. This is the simplest possible description, in contrast to a dust size distribution, as such distributions are not yet well constrained for protoplanetary disks. 
The assumption that dust of this size is well coupled to the gas is supported by observations, at least in some systems \citep{Pontoppidan2007, Sturm2023, Duchene2024, Villenave2024}.
We set a composition of carbon and silicate (pyroxene with 70\% magnesium and 30\% iron) with a mass ratio of carbon to pyroxene of $0.87/0.13$, and a porosity of 25\%, according to the DIANA standard dust model \citep{Woitke2016}.
We create the corresponding opacity using optool \citep{optool}. The resulting scattering and absorption opacities are shown in Figure~\ref{fig:chopping-opacity} in the appendix.

{To make sure that the qualitative results still hold for a different choice of dust grain sizes, we explored grain sizes of $a = 0.5\,\mu \mathrm{m}$ and $1\,\mu\mathrm{m}$, as well as a dust size distribution in Appendix~\ref{sec:grainsizes}. Overall, we find similar results for the lateral asymmetries and flux ratios in all of these models.}

The scattering phase functions in our set-up are very forward-peaked, especially for wavelengths up to a few micron. This can lead to very expensive simulations.
However, we can use a technique to save computation time, called chopping. In this technique, we chop off a few degrees of the scattering phase function and treat rays within that cone as if they were not interacting with the scattering particle.
This technique helps to avoid bright spots that appear in the radiative transfer models if not enough scattering photons are used.
We carefully checked the applicability of this technique in our case in Appendix~\ref{sec:chopping}.
For the Monte Carlo temperature calculations, we use $10^8$ photon packages. For the {near-/mid-infrared radiative transfer} images, we use $10^9$ scattering photon packages for all images in the wavelengths of $0.8\,\mathrm{\mu m}$, $2\,\mathrm{\mu m}$, and $21\,\mathrm{\mu m}$. We found that at $12\,\mathrm{\mu m}$, the same amount of photon packages leads to noisy images, which is why we use twice as much photon packages for this wavelength.

\begin{figure} [ht!]
  \centerline{\includegraphics[width=0.5\textwidth]{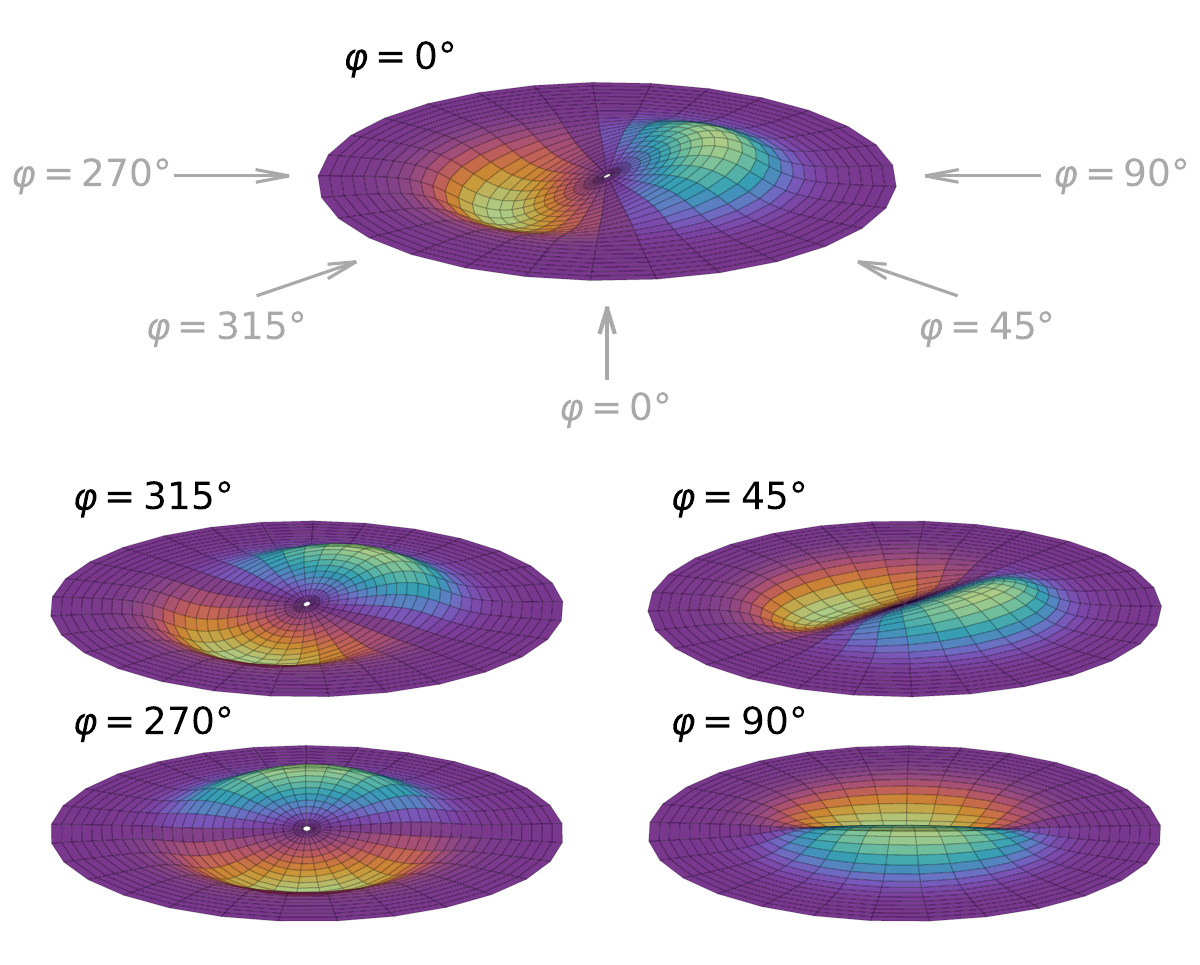}}
  \caption{\label{fig:phi-def} Definition of the azimuthal $\varphi$-angles. Colors represent the height above the $x$-$y$-plane: purple is in plane, blue-green above and red-yellow below. If the outer disk is viewed perfectly edge-on, the quadrant of $270\degree$–$0\degree$ is identical to the quadrant $0\degree$–$90\degree$, only flipped upside down. The sector $90\degree$–$270\degree$ is identical to the points of sector $270\degree$–$90\degree$ appropriately rotated and flipped. We chose this definition of $\varphi$ to be consistent with \citet{Villenave2024}.
  }
\end{figure}

Because warped disks are not axisymmetric, we need to synthetically observe the disk from different directions.
For that, we define an angle $\varphi$, so that $\varphi=0\degree$ looks at the warp tilt from the side, as illustrated in Figure~\ref{fig:phi-def}.
We can exploit some symmetries in the disk, as the disks in our models are untwisted.
Therefore, we only need to consider a quarter of the $\varphi$-space if the outer disk is viewed perfectly edge-on (inclination of exactly $i=90\degree$).
The remaining regime of $\varphi$ can then be achieved by appropriate rotations of the images.
In some models, we chose to look at the outer disk slightly inclined ($i=85\degree$).
In this case, we need to cover half of the azimuthal regime, as a part of the symmetry is broken.

\subsection{Analysis} \label{sec:methods-analysis}

{Images of edge-on protoplanetary disks in the mid- to near-infrared wavelength regime} typically show two nebulae separated by an elongated dark lane. 
In the presence of a tilted or warped inner region, the illumination of the upper disk surfaces does not follow spherical symmetry and depends on the viewing orientation. This can lead to changes in the flux ratio between the two nebulae and/or to lateral asymmetries.

\begin{figure} [ht!]
  \centerline{\includegraphics[width=0.5\textwidth]{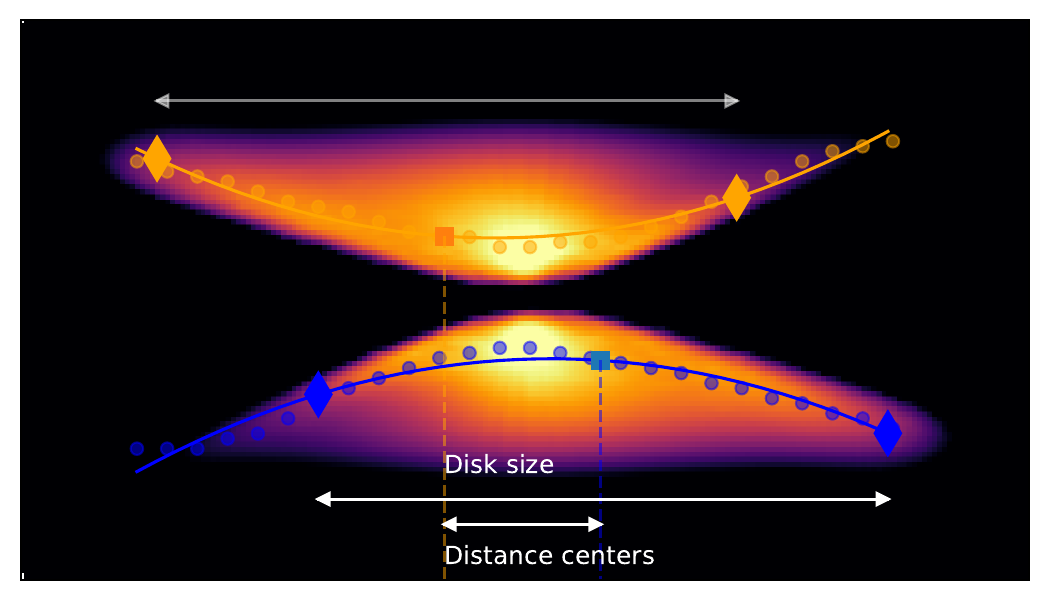}}
  \caption{Schematic representation of the spine fitting. The light circles show the location of the maxima after a high order polynomial fit of the minor axis profiles. The solid lines represent the final spines after fitting the maxima by a polynomial of order 2. The diamond represent the locations where the flux reaches 10\% with respect to the maximum of the spine, and the squares are the  center of the spine, i.e., the midpoint of these two points. The lateral asymmetry is the ratio between the distance of the centers to averaged disk size. 
  }
  \label{fig:exampleSpines} 
\end{figure}

\begin{figure*}[ht!]
\includegraphics[width=\textwidth]{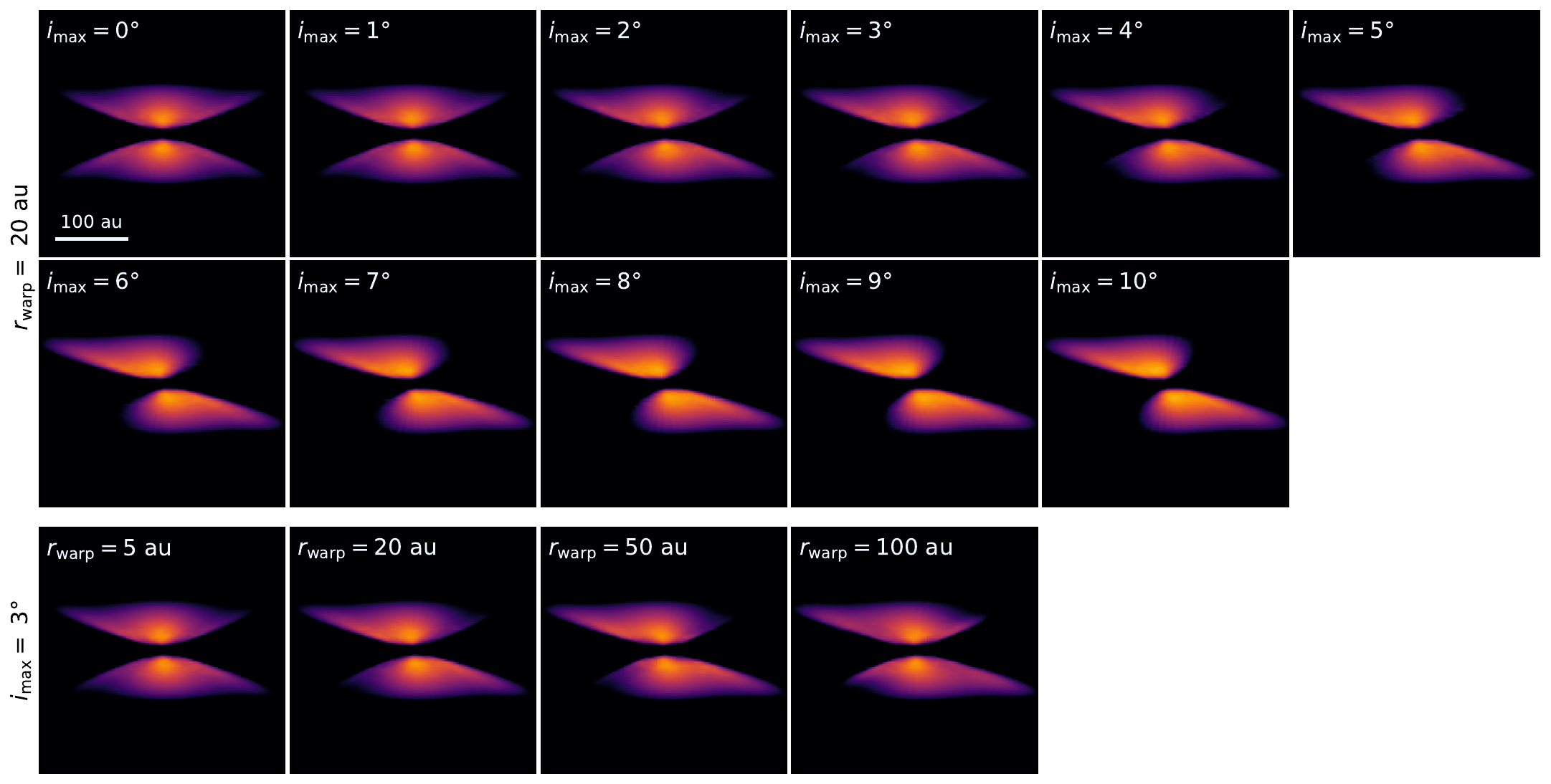}
     \caption{\label{fig:wampl} {Near-infrared radiative transfer} models of disks with different \revnew{amplitude warps, defined by} $i_\mathrm{max}$, observed at $\lambda~=~0.8\,\mathrm{\mu m}$ at $\varphi~=~0\degree$ (top two rows).
     The bottom row shows models with a varying warp location or in other words a varying size of the misaligned inner region. {For the color, we use a logarithmic scale.}
     }
\end{figure*}

To quantify the lateral asymmetry and flux ratio in relation to different warp parameters, we follow the methodology employed by \cite{Duchene2024} and following works on JWST observations. This method identifies a ``spine'' for each of the two scattering nebulae\footnote{We note that we call it spine due to the arc-like shape. It is not identical to the mathematical spline-fitting method.}. 
To do so, the method first estimates the location of the two maxima along the minor axis profiles for every location along the nebulae. The minor axis profiles, averaged to a resolution of 0.1'' or 0.2'' (for a disk model at a distance of $140\,\mathrm{pc}$), are fitted by a polynomial function of order 6 or 8, which is used to assess the location of the two maxima. Those are then fitted by a polynomial function of second order to determine the two spines. On both spines, we then identify the two points on each side of its maximum where the flux reaches 10\% with respect to the maximum. We define the midpoint of these two points as the center of the spine, and define the distance between them as the disk size. Finally, following \cite{Villenave2024}, we define the lateral asymmetry $\delta_\mathrm{spine}$ as the distance between the center of both spines in the direction along the nebulae's major axis, divided by the averaged disk size. 
The procedure described above is repeated multiple times for different radial extents on the scattering nebulae, between 0.9'' and 1.6'' with steps of 0.1'', and for other parameters (averaged resolution, polynomial order) in order to estimate an error on the measurement of the lateral asymmetry. A schematic representation of the spine fitting is shown in Fig.~\ref{fig:exampleSpines}.

In addition, independent of the spine fitting, we estimate the integrated flux ratio between the top and bottom nebulae. We do this by taking the ratio of the total flux included in the top nebula divided by the total flux of the bottom nebula.
We notice that this definition is the inverse of that of \cite{Villenave2024}.
Therefore the flux ratio in our unwarped models is greater than one for a disk inclined at $85\degree$, in contrast to theirs being smaller than one.

\section{Results} \label{sec:results}

Protoplanetary disks in {near-/mid-infrared} show, if they are viewed edge-on (at an observer inclination close to $90\degree$), two nebulae separated by a dark lane.
In the following section (Section~\ref{sec:results-single-wl}), we present the {near-infrared} images\footnote{We note that we do not show all images of all models we created, but a select sample. However, every dot in Figures~\ref{fig:spine_flux_strength},~\ref{fig:spine_flux_location}, and \ref{fig:flux_wavelength} is a full {near-/mid-infrared radiative transfer} simulation.} at $\lambda~=~0.8\,\mathrm{\mu m}$ for different warp morphologies, as well as our analysis of the lateral asymmetry and the flux ratio between the two nebulae.
In Section~\ref{sec:multiple-wavelengths}, we will discuss images at multiple wavelengths.
In Appendix~\ref{sec:unwarped}, we additionally show the models of an unwarped reference case, both perfectly edge-on and slightly inclined ($85\degree$). 

\subsection{{Near-infrared} images at a single wavelength}\label{sec:results-single-wl}

We present {near-infrared} images at $\lambda~=~0.8\,\mathrm{\mu m}$ of different warp amplitudes \revnew{(characterized by $i_\mathrm{max}$)} in Figure~\ref{fig:wampl}, top two rows. The bottom row in this figure shows models with different locations of the warp, or in other words different sizes of the misaligned inner region.
All disks in this figure are viewed perfectly edge-on and at an azimuthal angle of $\varphi=0\degree$, meaning that the warp tilt is viewed from the side (see Figure~\ref{fig:phi-def}).
For this viewing angle, the lateral asymmetry is maximal, as the shadowed regions of both nebulae are the most opposite.

\begin{figure} [ht!]
  \centerline{\includegraphics[width=0.5\textwidth]{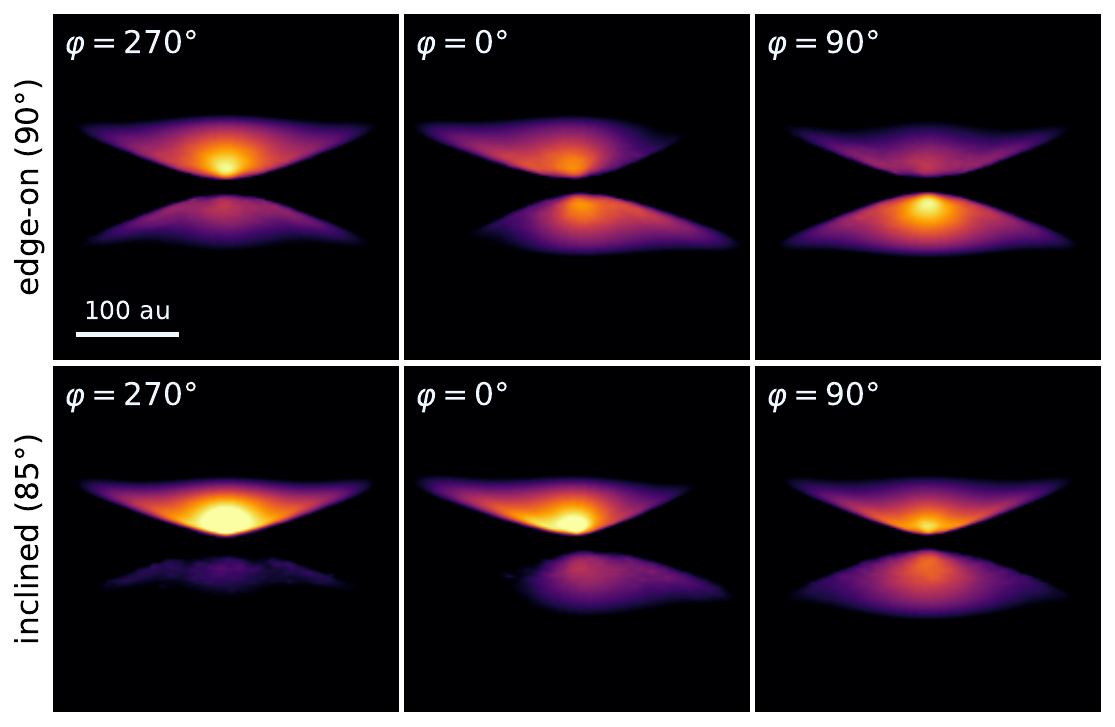}}
  \caption{\label{fig:tilt3-phi} {Near-infrared radiative transfer} models {($\lambda = 0.8\,\mu \mathrm{m}$)} of a disk with tilt $i_\mathrm{max}=3\degree$ observed from different azimuthal angles $\varphi$  {(color in log-scale)}. The top row shows a perfectly edge-on view, while the bottom row shows the disk observed at an inclination of $85\degree$. We note that for symmetry reasons, we do not need to simulate $\varphi~=~90\degree$. The image shown in this figure is the simulation of $\varphi~=~270\degree$ flipped upside down. }
\end{figure}

The orientation of the warped disk with respect to the observer has a large effect on the visibility of the asymmetry.
In Figure~\ref{fig:tilt3-phi}, we show how this picture changes for different azimuthal orientations in the model with a \revnew{misalignment between inner and outer disk} of $i_\mathrm{max}~=~3\degree$.
The top row shows the perfectly edge-on view (inclination of $90\degree$), while the bottom row presents a disk inclined by $85\degree$.
The figure shows that the lateral asymmetry vanishes for azimuthal viewing angles shifted by $90\degree$ from zero, as the misaligned inner region is viewed face-on in these cases (see Figure~\ref{fig:phi-def}, bottom row).

\begin{figure}
    \centering
    \includegraphics[width=\linewidth]{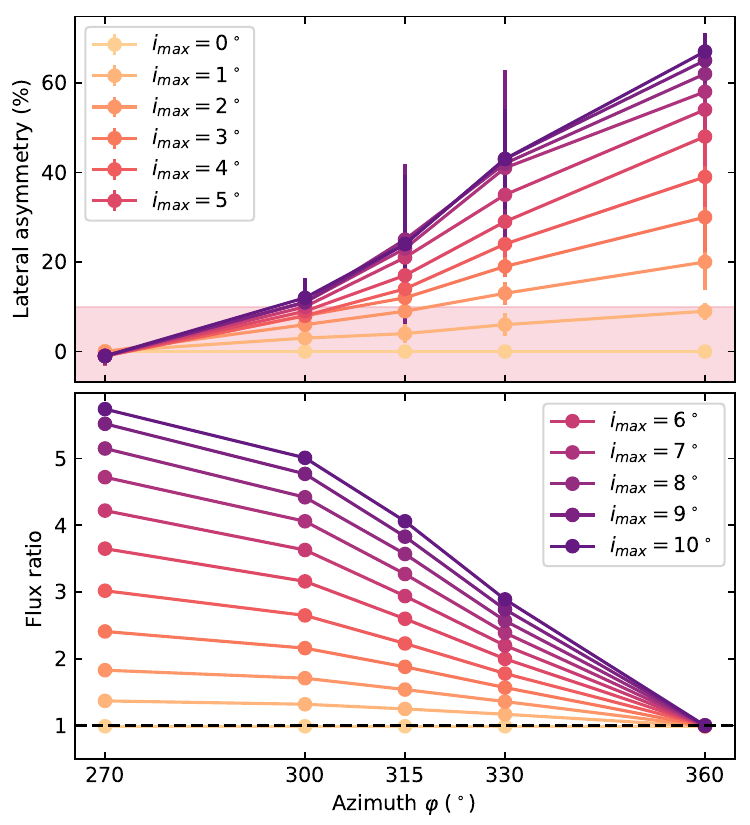}
    \caption{Lateral asymmetry (top) and flux ratios (bottom) for warped disk models of different warp tilts $i_\mathrm{max}$ {(at $\lambda~=~0.8\,\mathrm{\mu m}$)}. The outer disk is viewed perfectly edge-on (meaning an inclination of $90\degree$). For symmetry reasons, we only need to model a quarter of the azimuthal regime in this configuration. The curves can be extended to the remaining regime by flipping the curves appropriately. {The pink shaded region in the top panel corresponds to a lateral asymmetry of $<10\%$, which is what we assume not to be visible by eye.}}
    \label{fig:spine_flux_strength}
\end{figure}

\begin{figure}
    \centering
    \includegraphics[width=\linewidth]{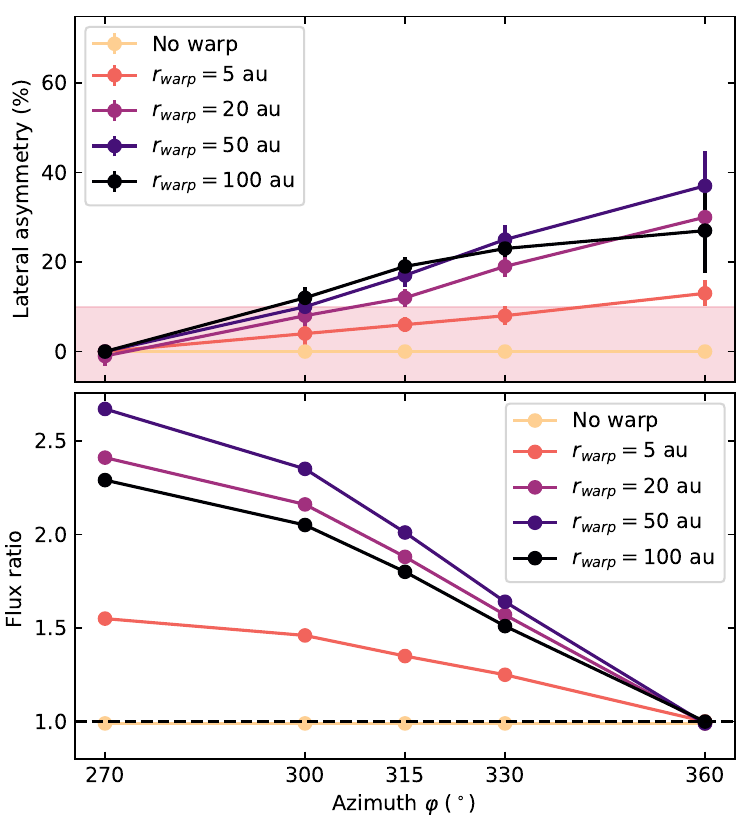}
    \caption{Same as Figure~\ref{fig:spine_flux_strength}, but for different warp locations $r_\mathrm{warp}$ for a {warp} amplitude \revnew{characterized by} $i_\mathrm{max}=3\degree$. Again due to symmetry reasons, the curves can be extended to the whole azimuthal regime. {As in Figure~\ref{fig:spine_flux_strength}, the pink shaded region highlights the lateral asymmetry we assume not to be visible by eye.}}
    \label{fig:spine_flux_location}
\end{figure}

To quantify the results of the {near-infrared} images, we evaluate the lateral asymmetry, as well as the flux ratio between the two nebulae as described in Section~\ref{sec:methods-analysis}.
Figures~\ref{fig:spine_flux_strength} and \ref{fig:spine_flux_location} present both quantities for our models with varying {warp amplitude} and warp location, respectively. In all those models, the outer disk is viewed perfectly edge-on.
By eye, we define a lateral asymmetry of $\geq 10\%$ to be visible and note that this is a somewhat arbitrary definition, but motivated by previously published work \citep[see][]{Villenave2024}.
{The pink shaded region in Figures~\ref{fig:spine_flux_strength} and \ref{fig:spine_flux_location} indicates the region below this visible threshold.}

We find that the stronger the warp (higher $i_\mathrm{max}$), the stronger the lateral asymmetry.
This result is expected, since stronger misalignments can cast larger shadows.
In Figure~\ref{fig:wampl}, we see that the lateral asymmetry of a warped disk can be visible in scattered light with a \revnew{misalignment} of only $2\degree$.
This means that even slight warps can in principle be observable.
However, we note that this is only the case for the best azimuthal viewing angle.
For viewing angles within $30\degree$ around $\varphi=270\degree$ (and $\varphi=90\degree$, respectively), even a stronger warp of $10\degree$ \revnew{misalignment} can not produce a significant lateral asymmetry (see Figure~\ref{fig:spine_flux_strength}).

The difference in flux between the two nebulae, on the other hand, is maximal for a viewing angle of $\varphi=270\degree$ and, for an edge-on view on the outer disk, equals one if the warp is viewed directly from the side ($\varphi=0\degree$).
In observations, the flux ratio by itself is not a unique indication of a warp or misalignment, since slightly inclined planar disks also exhibit a difference in flux between the two nebulae.
However, the variation of the flux ratio with wavelength can be a strong indicator of warps, which we will explore in Section~\ref{sec:multiple-wavelengths}.

When changing the location of the warp (bottom row in Figure~\ref{fig:wampl}), we can see a similar effect: the further out the warp, the stronger the asymmetry in the {near-infrared} images.
However, this only holds true up to a certain point, as a warp location of $100\,\mathrm{au}$ shows slightly less asymmetry than a warp location of $50\,\mathrm{au}$, see bottom row of Figure~\ref{fig:wampl} and Figure~\ref{fig:spine_flux_location}.
We suspect this to occur because of the way we set up the warp shape, as we scale width of the warp with the warp location. This description is the most natural, because it is coherent in logarithmic space.
However, considered linearly, the warp transition is wider for warps located further out, leading to a less steep warp curve where more light can pass.
This reduces the size of the shadowed region and causes less lateral asymmetry.

Additionally, we find the flux ratio for the case of $r_\mathrm{warp}~=~100\,\mathrm{au}$ to be smaller than for some warps closer in at all azimuthal angles.
This could again be linked to the set-up of the warp: Because the warp is close to the outer edge, the outer disk is small compared to the other models, which could influence both lateral asymmetry and flux ratio.

An interesting fact is that the image of the warped disk at $\varphi=270\degree$ with the outer disk viewed perfectly edge-on (top left image in Figure~\ref{fig:tilt3-phi}) looks very similar to an unwarped disk which is inclined with respect to the observer, see Figure~\ref{fig:unwarped} in Appendix~\ref{sec:unwarped}, bottom left panel.
For observations, this means that it can be difficult to distinguish between an inclined planar disk and a perfectly edge-on warped system.
We notice that in our models, the {near-infrared} image of the inclined planar disk is more strongly peaked toward the star.
However, this is unlikely to serve as distinction for observations, since the disk parameters are usually not known precisely.

\subsection{{Near-/mid-infrared radiative transfer} models in multiple wavelengths} \label{sec:multiple-wavelengths}

A few observations of protoplanetary disks show a swap of the brightest nebula with wavelength: in shorter wavelengths one nebula appears brighter, while the other nebula appears brighter in longer wavelengths.
This phenomenon is rare, so far it has only been observed in three disks: IRAS04302+2247, HH30, and 2MASS J16281370-2431391 (Flying Saucer).
The swap {does} not occur at the same wavelength for the observed disks.
In IRAS04302+2247, the swap occurs between $\lambda~=~12.8\,\mathrm{\mu m}$ and $21\,\mathrm{\mu m}$ \citep{Villenave2024}, while the switch for the Flying Saucer occurs between $\lambda~=~1.3\,\mathrm{\mu m}$ and $2.2\,\mathrm{\mu m}$ \citep{Grosso2003}. Finally, {\citet{Tazaki2025}} identified a change in the brightest nebula of HH30 between 7.7\,$\mu$m and 12.8\,$\mu$m.  {They mentioned intrinsic flux variability between epochs as a potential origin for a wavelength swap. However, with their observations being observed consecutively within 3h, they ruled out this possibility for this system. On the other hand,}  
\cite{Villenave2024} were able to recreate this phenomenon in a broken disk with a misaligned inner region.
In this work, we aim to investigate if a continuously warped disk exhibits a similar behavior.

\begin{figure} [ht!]
  \centerline{\includegraphics[width=0.5\textwidth]{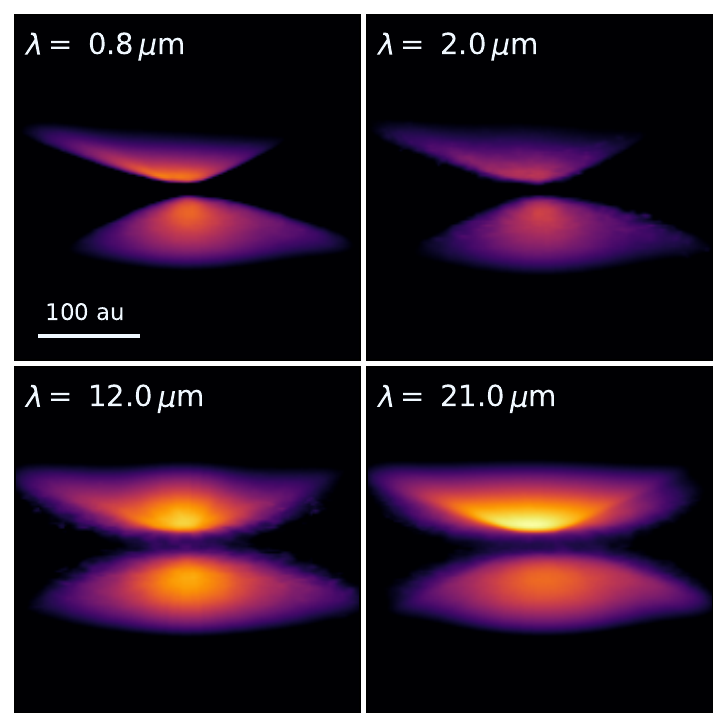}}
  \caption{\label{fig:tilt10-lambda} {Near-/mid-infrared radiative transfer} models of a disk with tilt $i_\mathrm{max}=10\degree$ at $\varphi=60\degree$ with an inclination of $85\degree$ observed in different wavelengths {(color in log-scale)}. Considering the integrated flux, the bottom nebula is brighter than the top at shorter wavelengths, but at $\lambda~=~21\,\mathrm{\mu m}$, the top nebula is brighter.}
\end{figure}

For that, we present radiative transfer simulations in four different wavelengths: $\lambda~=~0.8\,\mathrm{\mu m}$, $2\,\mathrm{\mu m}$, $12\,\mathrm{\mu m}$, and $21\,\mathrm{\mu m}$.
We test the multi-wavelength behavior for two different warp tilts $i_\mathrm{max}~=~3\degree$ and $10\degree$.
In these models, the outer disk is viewed at an inclination of $85\degree$, because the phenomenon only occurs for slightly inclined disks, as discussed later in this section.
We indeed find a swap in brightest nebula in specific cases.
Figure~\ref{fig:tilt10-lambda} shows an example of such a case, in especially, a warp tilt of $10\degree$ and an azimuthal viewing angle of $\varphi~=~60\degree$.
For the shortest wavelength, $\lambda~=~0.8\,\mathrm{\mu m}$, the bottom nebula is brighter, while for $\lambda~=~21\,\mathrm{\mu m}$, the top nebula clearly appears brighter.

The phenomenon of the switch between brightest nebula can be explained by the shadow from the misaligned inner region.
First, we need to understand that in a general disk that is slightly inclined with respect to the line of sight, one of the nebulae appears brighter than the other.
{This is because forward scattering is dominant over backward scattering \citep[see e.g.][their Figure~2]{Benisty_2023}, which makes the nebula closest to the observer brighter. }
Now, an inclined warped disk could be oriented in such a way that this brighter nebula is actually shadowed by the misalignment. In this case, the nebula actually appears darker.
However, in longer wavelengths, the disk could be able to emit more thermal radiation and therefore the shadow would be less pronounced. Thus, the nebula can again appear brighter.
To test the hypothesis, we performed additional radiative transfer simulations with the scattering switched off (option \texttt{scattering\_mode\_max=0} in RADMC-3D).
We indeed find almost no emission for $\lambda~=~0.8\,\mathrm{\mu m}$ and $2\,\mathrm{\mu m}$, indicating that the light in the scattering simulations indeed results purely from scattered light.
In contrast, we {find thermal emission} in these test simulations in the wavelengths $\lambda~=~12\,\mathrm{\mu m}$ and $21\,\mathrm{\mu m}$, {where it is more significant for the latter.
However, we want to point out that the scattered light still dominates the appearance of the disks at both of these wavelengths.
In summary, the fact that the thermal emission becomes more significant for longer wavelengths means that the shadows in the scattered light become less pronounced, which enables the swap in brightest nebula with wavelength.}

\begin{figure}[ht!]
    \centering
    \includegraphics[width=\linewidth]{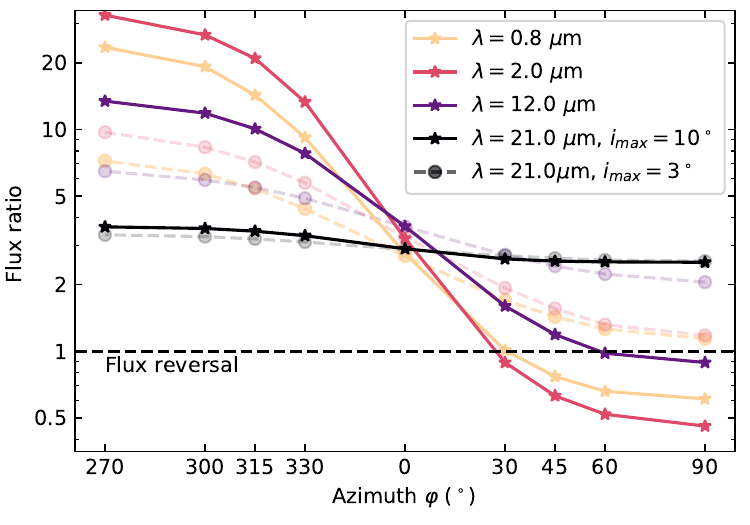}
    \caption{Flux ratios for models of warped disks with a warp tilt of $10\degree$ (solid lines) and of $3\degree$ (light dashed lines) in four different wavelengths.  The outer disk is oriented with an inclination of $85\degree$. In this case, we need to calculate the models for half of the $\varphi$-regime, the other half is symmetric. The dashed black line indicates the line where the flux ratio equals one (where both nebulae have the same brightness). We note that the $y$-axis is plotted in log-scale.}
    \label{fig:flux_wavelength}
\end{figure}

Evaluating the flux ratios quantitatively in Figure~\ref{fig:flux_wavelength}, we see that for a $\varphi$-regime of between $30\degree$ and $90\degree$ (and due to symmetry also up to $120\degree$), a flux ratio swap between wavelengths can be possible.
This only occurs for the stronger warp of $i_\mathrm{max}~=~10\degree$ in our models. For the moderate warp of $i_\mathrm{max}~=~3\degree$, the variation of the flux ratio with azimuthal angle is less strong at all wavelengths, so that the ratio never drops below one.

Concentrating on the disk containing a stronger warp, we see that the shape of the flux ratio curve differs with wavelength. Generally, for shorter wavelength the curve is steeper and there is a stronger variation of the flux ratios with azimuth.
Interestingly, a wavelength of $\lambda~=~2\,\mathrm{\mu m}$ shows more variation than $\lambda~=~0.8\,\mathrm{\mu m}$.
This could be due to a difference in the chopped scattering opacity and consequently a difference in albedo (see Figure~\ref{fig:chopping-opacity}).
The flux ratio curve for $\lambda~=~21\,\mathrm{\mu m}$ is shallowest, leading to a flux ratio always above one.
At the shorter wavelengths, the flux ratio drops below one for orientations where the misalignment is shadowing the otherwise brighter nebula.
We see that this swap of brightest nebula can occur at different wavelengths, depending on the azimuthal orientation of the disk.
Recall that we evaluate the flux ratio by using the integrated flux in our models.
We additionally checked the flux ratios when considering only the peak intensity and found that the curves are qualitatively the same, but slightly noisier and shifted, so that only the curve of $\lambda~=~2\,\mathrm{\mu m}$ drops below one.
However, the swap still occurs and with slightly different parameters, such as a different inclination of the outer disk, a similar result can be achieved in peak intensity flux ratios.

In our models, we notice that the azimuthal curve at $21\,\mathrm{\mu m}$ follows the same qualitative shape as the shorter wavelengths, with their maximum at $\varphi~=~270\degree$ and their minimum at $\varphi~=~90\degree$ (see Figure~\ref{fig:flux_wavelength}).
On the other hand, \cite{Villenave2024} find that the $21\,\mathrm{\mu m}$ curve is flipped with respect to the shorter wavelengths.
This difference could lie in the different opacities and geometry of the models.
Specifically, \cite{Villenave2024} investigated a broken disk, while we investigate a continuously warped disk.
In a broken disk, the inner edge of the outer disk and the outer edge of the inner disk could thermally radiate onto the outer disk in longer wavelengths and therefore leading to the inversion of the curve.
Because in the models of this work, there are no gap edges, we expect this thermal emission to be less strong.

At this point, we want to note that we expect the phenomenon of the swap in brightest nebula with wavelength to occur rarely, because a very specific orientation of the observed disk is required.
In especially, the outer disk needs to be slightly inclined with respect to the observer, and the azimuthal viewing orientation of the system is restricted to a specific regime with respect to the orientation of the misalignment.
This expectation connects to the fact that as of now, only three disks have been observed to exhibit this phenomenon.
However, if observed, a change in the brightest nebula with wavelength can be a strong indicator of a misalignment between inner and outer disk.

\subsection{Caveats} \label{sec:caveats}

The findings in this work provide valuable insight, although they should be considered within their limitations, which we discuss in this section.

Most importantly, the results significantly depend on the disk model.
The vertical thickness and flaring of the disk influence the extent of the shadows.
In thinner disks with a lower aspect ratio, we expect a stronger lateral asymmetry, because a larger part of the outer disk's surface can be hidden behind the misalignment. In thicker disks, on the other hand, the outer disk can be puffed up enough to bulge out of the shadow.
A similar argument applies when considering differently flared disks, which occurs for different temperature structures. With more flaring, the surface can bulge out more, leading to less lateral asymmetry.
In the same line of arguments, the size of the disk, in especially with respect to the location of the warp, can also influence the lateral asymmetry.
Additionally, the warp parametrization has a strong impact on the shadowed region. As it is not well constrained yet, we studied one of simplest warp shapes in this work: a warp that is only tilted in one direction.
In general, there can be more complicated shapes, for example twisted disks or multiply warped/twisted disks, leading to more complex shadow signatures.
Since the overall parameter space for disk and warp shapes is extremely large, a further exploration would go beyond the scope of this work.

Another factor influencing the results of the {near-/mid-infrared} images is the scattering opacity, which depends on the dust properties and total dust mass \citep{Watson_2007}.
A higher dust mass would lead to a higher opacity and we therefore expect the dark lane between the nebulae to be wider, possibly influencing the lateral asymmetry.
Different dust sizes or a size distribution can also strongly affect the opacity.
In our models, we only consider small dust grains that are perfectly coupled with the gas.
Although it is likely that larger dust particles are present in protoplanetary disks, we do not expect them to have a strong effect, as they are likely settled to the midplane \citep[e.g.][]{Villenave_2020}.
We additionally recall our use of the so-called chopping mechanism of $2\degree$  to calculate the scattering matrix due to the forward peaked phase functions at small wavelengths.
Although we thoroughly tested the applicability of this mechanism to our case in Appendix~\ref{sec:chopping}, it remains an approximation.
We notice that we slightly overestimate the total flux by a few percent when using the chopping mechanism. 
Since we are mainly interested in the lateral asymmetry and flux ratios, this does not have a large effect on the results in this work.

We also note that the {near-/mid-infrared} images presented in this study are not convolved to a specific instrument resolution, in order to keep this study general as convolution may depend on specific telescope settings.
As the images are fairly smooth, we do not expect a convolution to significantly change the results.

Lastly, we mention that our definition of the lateral asymmetry is somewhat arbitrary. We lean the definition on previously published work, but it can be defined differently, which could have an impact on the results.
Our fitting method strongly depends on the parameters chosen, namely the radial range of the spines, the radial and vertical averaging of the minor axis profiles, the order of the polynomial function fitting the minor axis profiles, the assumed location for the center of the dark lane. In this work we explored how varying these parameters will impact the results, reaching to error bars between a few percents to tens of percents (see Fig.~\ref{fig:spine_flux_strength} and Fig.~\ref{fig:spine_flux_location}). However, due to the limited parameter space explored, the relative comparison between models are likely more robust than the absolute values of the lateral asymmetries.

\section{Occurrence of warps in protoplanetary disks} \label{sec:jets}

In Section~\ref{sec:results}, we investigated the level of lateral asymmetry that can be expected for different warp {amplitudes} and locations, under our fiducial assumptions for dust opacities and warp geometry. Here, we aim to link these predictions to direct observations of edge-on protoplanetary disks. We compare the outer disk and jet orientation of four previously published edge-on disks observed at scattered light wavelengths. Our goal is to look for misalignment between the very inner regions of the disks, where the jet is launched, and the outer disk seen in scattered light or at millimeter wavelengths. Then, we compare apparent misalignments with the observed lateral asymmetry to assess the potential geometry of the systems. 

\begin{figure}[ht!]
    \centering
    \includegraphics[width=\linewidth]{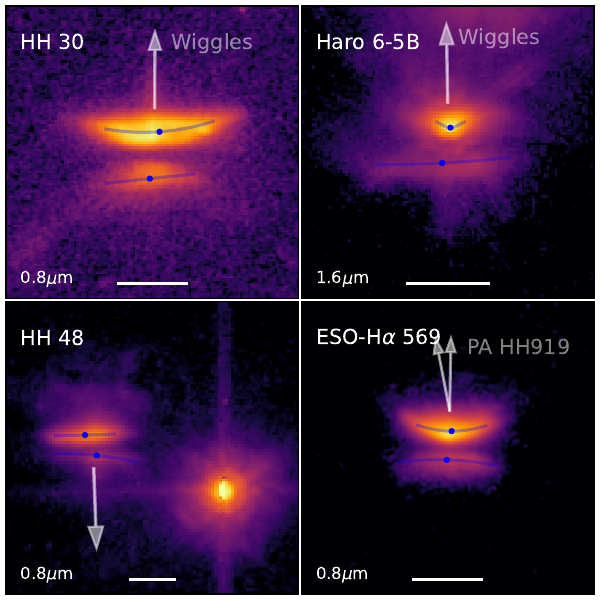}
    \caption{Scattered light observations of four edge-on disks {in log-scale}, where we estimate the level of lateral asymmetry (blue spines) and indicate the estimated direction of the jet (white lines). {A 0.5\arcsec\ horizontal line is indicated in the bottom left corner of the images.}}
    \label{fig:misaligned}
\end{figure}

We selected the disks based on the recent study of \cite{Villenave2024}, who found that  15 out of the 20 systems that they analyzed display strong lateral asymmetries. Seven sources of this sample display clear jet features,
among which we identified four sources with a published value for the position angle (PA) of their jets from high angular resolution observations in the literature (with the other three disks, HV Tau C, Tau042021, and ESO H$\alpha$-574 having a jet whose PA is not explicitly characterized in the literature; \citealt{Stapelfeldt_2003, Duchene_2014, Duchene2024}). We report the values for the jet position angle in Table~\ref{tab:jetcomparison}. {We note however that the published jet PA of HH48, reported by \citet{Stapelfeldt_2014}, suffered from image distortion in the HST Advanced Camera instrument (K. Stapelfeldt, priv. comm.). In Table~\ref{tab:jetcomparison}, we report our visual estimate of the jet PA on the distortion corrected image, leaving a more precise measurement for future work.} All disks have been observed and well characterized at millimeter (with ALMA) or scattered light wavelengths (with HST), so we additionally report the outer disk position angle in Table~\ref{tab:jetcomparison}. If no misalignment is present, we expect the position angle of the jet to be exactly perpendicular to that of the disk's major axis. The scattered light images together with the corresponding jet orientation are displayed in Fig.~\ref{fig:misaligned}.

\begin{table}[ht!] 
    \centering
    \caption{Comparison between outer disk position angle ($\text{PA}_{out}$) and jet position angle ($\text{PA}_{jet}$)}
    \begin{tabularx}{0.5\textwidth}{p{1.3cm}p{1.3cm}p{1.3cm}p{1.5cm}p{2cm}}
    \hline \hline
    Disk & HH30 & HH48 & Haro6-5B &    ESO H$\alpha$-569\\
    \hline
    PA$_{out}$ ($^\circ$) & 121$^{(1)}$ & 75$^{(1)}$ &  145$^{(1)}$ & 155$^{(9)}$\\
    PA$_{jet}$ ($^\circ$) & 31$^{(2, 3, 4)}$ &$\leq${13}$^{(5)}$ &  56$^{(6, 7, 8)}$  & $60-75^{(10)}$ \\
    $\vartheta$ ($^\circ$) & 0 & {$0-2$} &  1& $0-10$ \\
     $\delta_\mathrm{spine}$  (\%) &$5-9$ & $8-15$ & $10-36$ & $3-5$ \\ 
    Notes jet/disk & Bending, wiggling &  Perturbed nebulae& Wiggling &Assuming jet is HH919 filament\\
    \hline
    \end{tabularx}
    \tablefoot{$\vartheta =  |\left[\text{PA}_{out} -90\right] - \text{PA}_{jet}|$ characterizes the misalignment between the jet (inner disk) and the outer disk. The lateral asymmetry $\delta_\mathrm{spine}$ is fitted using the same method as in the models (see Sect.~\ref{sec:methods-analysis}).}
    \tablebib{
    (1)~\citet{Villenave_2020}; (2) \citet{Burrows1996}; (3) \citet{Anglada_2007}; (4) \citet{Estalella_2012};
    (5) this work; (6) \citet{Eisloffel_1998}; (7) \citet{Krist_1998}; (8) \citet{Woitas_2002};
    (9) \citet{Wolff_2017}; (10) \citet{Bally_2006}
    }
    \label{tab:jetcomparison}
\end{table}

All four systems show hints towards small levels of misalignment of their inner disk.
In the case of HH48, we find a misalignment between jet and {disk bottom nebula of about 2$^\circ$, while no misalignment with the top nebula is identified.} {Indeed, HH48 displays an apparent misalignment between its top and bottom nebulae, hinting toward a distorted morphology.
A somewhat similar misalignment can also be found in some of our models due to illumination effects, such as in Figure~\ref{fig:tilt10-lambda}, top two panels. However, for a good comparison, the current observational constraints on the dust and gas morphology of HH48 are not sufficient.}
In ESO H$\alpha$-569, the jet visible in the scattered light image does not have a published PA, but is thought to be launching the HH919 jet, which could be consistent with either being aligned, or misaligned to the outer disk by up to 10$^\circ$.
Finally, even though the jet and disk of HH30 and Haro6-5B are consistent with no misalignment, the jet of both sources wiggles at large distances. This can be explained in case of varying orientation of the accreted material, suggesting a precession of the inner disk~(\citealt{Terquem_1999}; of order of $\sim1^\circ$ for HH30, \citealt{Louvet_2018}). {Alternatively, interaction with a non uniform environment may lead jets to bend. Regular wiggles as observed in these systems are however not necessarily expected in that case.}

Following the methodology previously described (Sect.~\ref{sec:methods-analysis}, \citealt{Villenave2024}), we estimated the lateral asymmetry in these disks, as represented on Figure~\ref{fig:misaligned} and indicated in Table~\ref{tab:jetcomparison}. The strongest lateral asymmetry is found around Haro6-5B ($10-30\%$), while ESO H$\alpha$-569 is barely asymmetric ($3-5\%$). HH48 and HH30 show intermediate levels of asymmetries of order 10\%.

HH48 shows
{the most convincing case for a distorted morphology and hints to a misalignment between the jet and the disk plane of up to 2$^\circ$. Further, the disk}
presents a significant lateral asymmetry. Within the modeling presented in Sect.~\ref{sec:results}, we found that lateral asymmetry around 10\% could be obtained for a \revnew{warped disk with a misalignment between inner and outer disk} of $i_{max}=1-2^\circ$ if viewed with the optimal orientation ($\varphi\sim360^\circ$), or for stronger warps, up to $i_{max}=10^\circ$, for less favorable viewing angles ($\varphi\rightarrow270^\circ$), which would be consistent with the apparent jet/disk misalignment observed. 
Both the significant lateral asymmetry and the locally aligned but wiggling jets of HH30 and Haro6-5B suggest that these systems could also possess warps, which are, however, weak or not viewed in a favorable orientation. Lastly, ESO~H$\alpha$~569 could be consistent with either no misalignment or an unfavorable warp orientation.

Finally, we also mention that {a local change in the disk scale height, leading to asymmetric shadowing of the outer disk, or}  variation in the stellar illumination~\citep[e.g., spots;][]{Stapelfeldt_1999} could also lead to lateral asymmetries. Dedicated imaging campaigns would be needed to determine the timescale of the variability of the lateral asymmetry and disentangle {the} mechanisms.

\section{Conclusions} \label{sec:conclusion}

In this work, we present various {near-/mid-infrared} images from radiative transfer simulations of edge-on protoplanetary disks containing a warp.
Edge-on disks in scattered light typically show two nebulae separated by a dark lane. These nebulae correspond to the two surfaces of the disk, whereas the dark lane corresponds to the optically thick midplane.
Warped disks contain a misaligned inner region casting a shadow onto the outer disk, leading to visible asymmetries between the two nebulae.
We quantify this asymmetry by fitting a spine curve for each nebula individually and call this relative distance in between the nebula centers ``lateral asymmetry''.
We investigate the visibility of the lateral asymmetry for different warp parameters, especially the warp {amplitude} and the location of the steepest warp transition, or in other words different sizes of the misaligned inner region.
Additionally, we evaluate the flux ratio between the two nebulae.
Because warped disks are not axisymmetric, we performed radiative transfer simulations for images viewed from different azimuthal angles.
Most of our simulations are performed at a wavelength of $\lambda~=~0.8\,\mathrm{\mu m}$, but we extend the simulations to $\lambda~=~2\,\mathrm{\mu m}$, $12\,\mathrm{\mu m}$, and $21\,\mathrm{\mu m}$ in specific cases.

We find that a warp can be visible in edge-on disks for warp tilts as low as $i_\mathrm{max}=2\degree$ in optimal azimuthal viewing angles.
The larger the warp tilt, the stronger the lateral asymmetry.
Additionally, the location of the warp within the disk influences the asymmetry as well. If the warp is located further out, the asymmetry is stronger.
This means that in edge-on disks, warps close to the inner edge of the disk are harder to detect.
However, our models show a point where the asymmetry decreases for even further warp locations. We suspect this to occur due to the warp transition width which is scaling with radius. This leads to a shallower warp curve, and consequently more light can pass, causing smaller shadows and therefore less lateral asymmetry.

The flux ratio between the two nebulae can also indicate a warp in specific cases.
In unwarped and perfectly edge-on disk, the two nebulae have the same brightness. Due to the shadows in warped disks, the flux ratio can vary depending on the azimuthal angle.
By itself, the flux ratio is not a strong indicator of a warp, since the relative brightness of the two nebulae also varies in an unwarped disk when it is not oriented perfectly edge-on.
Nevertheless, we find that in some rare occasions, the flux ratio can be an indication of warps in observations after all, when observing the disk in multiple wavelengths.
For this scenario, the outer edge of the warped disk should be slightly inclined (not perfectly edge-on; we chose an inclination of $85\degree$ in our models).
If the disk is additionally oriented azimuthally, such that the nebula that would be brighter if unwarped is now shadowed by the misalignment, this nebula appears darker in {the near-infrared}.
At slightly longer wavelength ($12-21\,\mathrm{\mu m}$ in our models), however, the disk's thermal emission is large enough so that this nebula can appear brighter, as the shadow is less prominent, {leading to a swap in brightest nebula with wavelength.}
We confirm this swap in brightest nebula with wavelength if the warp is strong enough, in our models $i_\mathrm{max}~=~10\degree$.
For small warps ($i_\mathrm{max}=3\degree$), we can not reproduce this switch even at the best viewing angle.

In observations, such a swap is found in three edge-on disks so far: IRAS04302+2247, HH30, and Flying Saucer.
As expected, this phenomenon is rare, as it requires a very specific orientation of the disk with respect to the observer.
However, when it occurs, it is a strong indicator for a misalignment.
In observations, the inclination of an observed disk is often inferred from the flux ratio. If the observed disk contains a misalignment, it could lead to a mismatch of inclinations inferred from different wavelengths. Such a mismatch is indeed found for example in {\cite{Tazaki2025}.}
{We want to point out that observed disk inclinations usually have large uncertainties. However, if a swap with wavelength is observed, it can be a good indication that the disk is inclined and warped.}

Additionally, we analyzed the relative position angle between optical jets and the outer disk plane in four previously published edge-on protoplanetary disks.
We specifically targeted disks showing some level of lateral asymmetry in previous scattered light observations.
In our investigation, we find that all systems show hints to slight misalignments.
We use our models to make a rough estimation of the {amplitude} of misalignment in those disks. For HH48, we find that depending on azimuthal orientation, a warp between $1-10\degree$ could be present, consistent with the measurement of jet and {bottom disk surface misalignment ($\lesssim2^\circ$).}
In the other systems, the misalignments of a few degrees could be consistent with the apparent wiggling of the jet and the lateral asymmetries in the nebulae. {In summary, the weak or lack of misalignments between the jets and outer disks could indicate that warps in protoplanetary disks are typically only of a few degrees. However, more observed systems are needed in order to draw strong conclusions. }

Lateral asymmetries in edge-on observations can point to the presence of a warp, and with additional information about the vertical structure of the disk, it can be possible to make estimations about the warp parameters.
As always, these results must be interpreted within their limitations, which we elaborated in Section~\ref{sec:caveats}.
Although it is hard to infer the exact warp parameters from observations at this point, this study takes the first step in quantifying the asymmetry and trying to match the warp strength to the observed asymmetry.
However, the parameter space of different warp shapes is too large to cover in one single work, and we therefore have to leave further explorations for future studies.

\begin{acknowledgements}
We thank Cornelis Dullemond, Giovanni Rosotti, {Karl Stapelfeldt,} Stefano Facchini, and Giuseppe Lodato for their valuable insight and support for this project.
\revnew{We sincerely appreciate the thoughtful and detailed feedback from the referee, which has enhanced the clarity and depth of our work.}
We also want to thank Philipp Weber and Thomas Rometsch for their numerical and contentual help.
We also thank Ryo Tazaki, François Ménard, Gaspard Duchêne and the participants of the edge-on disk workshop (founded by the ERC grant agreement No. 101053020, project Dust2Planets, PI F. Ménard) which provided valuable feedback to this project.
We remember the legacy of Willy Kley, who passed away in 2021.
CK acknowledges funding from the DFG research group FOR 2634 “Planet Formation Witnesses and Probes: Transition Disks” under grant DU 414/23-2 and KL 650/29-1, 650/29-2, 650/30-1.
MV and CK acknowledge support from the European Union (ERC Starting Grant DiscEvol, project number 101039651, PI: G. Rosotti). {CK acknowledges support from Fondazione Cariplo, grant No. 2022-1217. Views and opinions expressed are, however, those of the author(s) only and do not necessarily reflect those of the European Union or the European Research Council. Neither the European Union nor the granting authority can be held responsible for them.}
Additionally, we acknowledge support by the High Performance and Cloud Computing Group at the Zentrum für Datenverarbeitung of the University of Tübingen, the state of Baden-Württemberg through bwHPC and the German Research Foundation (DFG) through grant INST 37/935-1 FUGG.
Plots in this paper were made with the Python \citep{Python3} library \texttt{matplotlib} \citep{matplotlib}. We acknowledge also the usage of \texttt{numpy} \citep{NumPy}, \texttt{astropy} \citep{AstroPy}, and \texttt{scipy} \citep{SciPy}.

\end{acknowledgements}

\bibliographystyle{aa}
\bibliography{edgeon}

\begin{appendix} 

\section{On the forward scattering peak} \label{sec:chopping}
For this work, we needed to perform a large amount of simulation runs, as each viewing angle $\varphi$ in each wavelength requires its own {radiative transfer} simulation.
In order to keep the computation time feasible for this study, each simulations should not take too long to run.
One of the main factors for computation time in {radiative transfer} simulations is the amount of scattering photon packages used.
The more photon packages are used, the closer the result gets to reality, but also the longer the simulation takes.

\begin{figure} [ht!]
  \centerline{\includegraphics[width=0.5\textwidth]{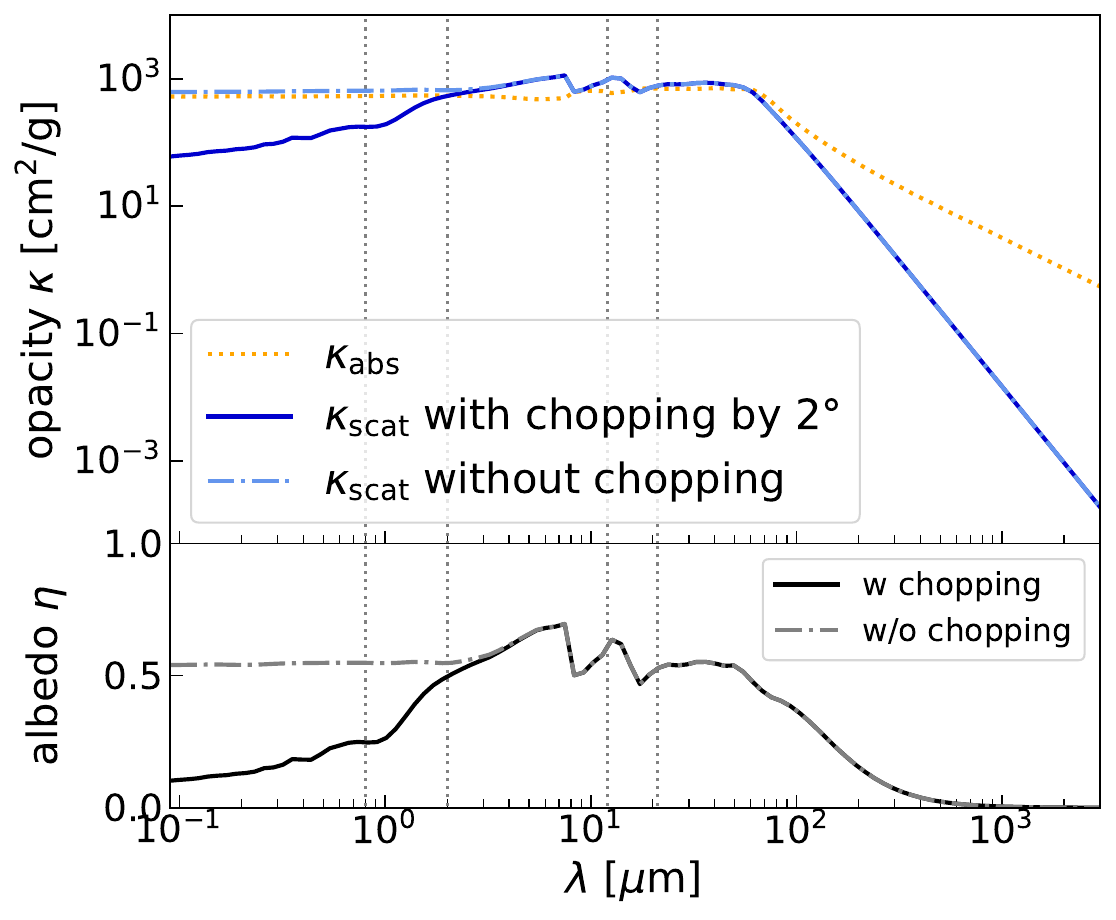}}
  \caption{\label{fig:chopping-opacity} Scattering opacity (mass-weighted) with chopping of $2\degree$ (solid) and without chopping (dash-dotted) for the DIANA standard model, as used in this work.
  The mass-weighted absorption opacity is indicated with the orange dotted line.
  The lower panel shows the albedo $\eta = \kappa_\mathrm{scat} / (\kappa_\mathrm{abs} + \kappa_\mathrm{scat})$.   
  The gray dotted lines indicate the wavelengths investigated in this work.
  }
\end{figure}

A commonly used technique to get away with fewer photon packages, but still end up with a smooth result, is called angular chopping \citep[see e.g. documentation of \texttt{optool},][]{optool}.
This technique makes use of the fact that light, which is scattered within only a few degrees around the forward scattering direction, can be approximated as unscattered.
This means that these photon packages, that would be scattered within this forward peaked regime, can be treated as if they did not interact with the scattering particle.
Internally, \texttt{optool} uses the chopping angle (option \texttt{-chop} in \texttt{optool}) to set the values of the scattering matrix within the cone of the set angle to the value at the cone edge.
The tool compensates for the chopping in order to ensure energy conservation by adjusting the scattering cross-section, which leads to a slightly altered opacity for wavelengths with a significant forward peak in the phase function.
Figure~\ref{fig:chopping-opacity} shows the comparison of the scattering opacity with and without chopping.

Clearly, chopping is especially important if the scattering phase function is extremely forward peaked – meaning that most photons are scattered within only a few degrees from the original direction.
Indeed, for short wavelengths, where the forward scattering peak in the phase function is extreme, Figure~\ref{fig:chopping-opacity} shows that the chopping mechanism reduces the opacity in this regime, while it does not influence the longer wavelengths, where the phase function is not as strongly forward-peaked.
Without the chopping mechanism, radiative transfer codes can have difficulty to deal with such an extreme forward peak, leading to bright spots in the resulting image, if too few scattering photons are used.
These bright spots can indeed be avoided by using an enormous amount of scattering photons.
However, this would lead to long computation times\footnote{On our machine with our settings, an image simulation with RADMC-3D using $10^9$ scattering photons takes less than one hour, while a simulation with $3 \cdot 10^{10}$ scattering photons takes more than a day. This would be unfeasible for a project like this work.}.

\begin{figure} [ht!]
  \centerline{\includegraphics[width=0.5\textwidth]{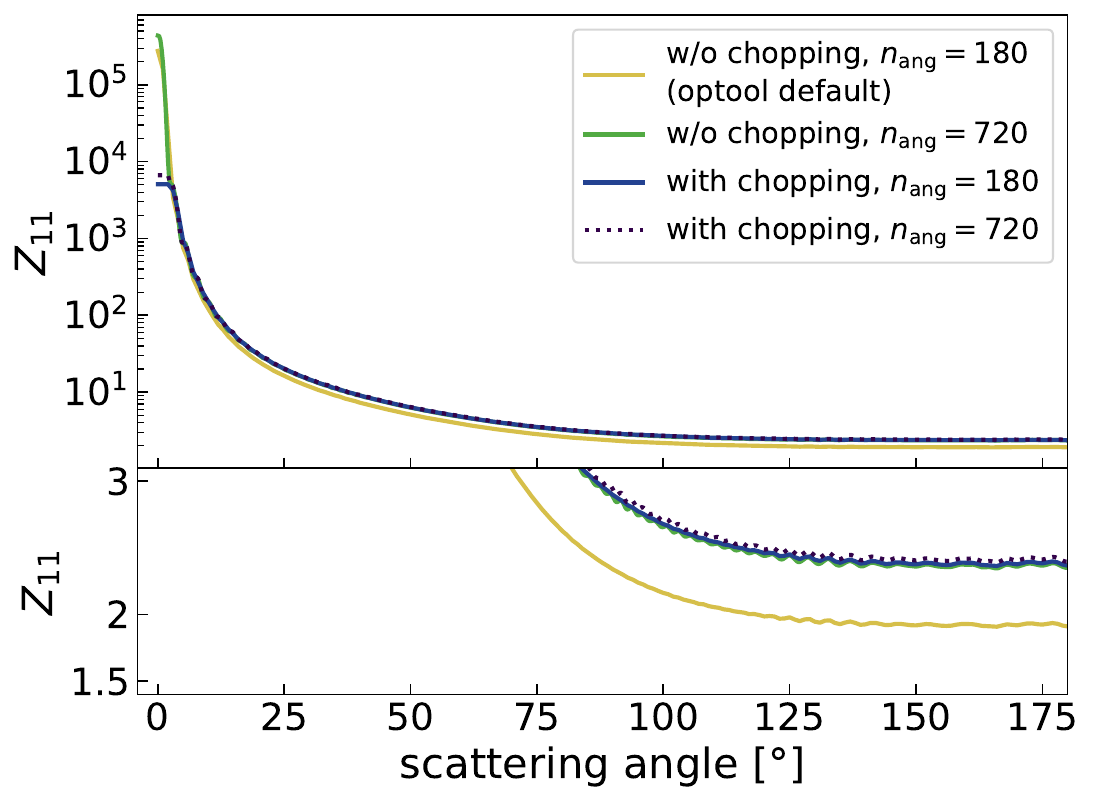}}
  \caption{\label{fig:phase-function} Phase function, i.e. $Z_{11}$-component of the scattering matrices, for the DIANA standard model in the wavelength $\lambda=\,0.8\mathrm{\mu m}$ without the chopping mechanism and with chopping by $2\degree$, using varying internal resolutions ($n_\mathrm{ang}$) in \texttt{optool}.  The bottom panel shows the same curves as the top panel, but with the $y$-axis linear and stretched so that the difference for larger scattering angles becomes visible. Except for the cut by $2\degree$ at the forward peak, the phase function is nearly the same for the case without chopping and high internal resolution (green line) and with chopping for the default internal resolution (blue solid line).}
\end{figure}

By looking at the phase function for our dust set up, we see that at the wavelength of $0.8\,\mathrm{\mu m}$, the phase function indeed is extremely forward peaked.
In Figure~\ref{fig:phase-function}, we plot the $Z_{11}$-component of the scattering matrix, whose integral determines the total scattering cross-section $\sigma_\mathrm{scat}$ with \citep[e.g.][]{radmc3d}
\begin{equation}
    \sigma_\mathrm{scat} = \int_0^\pi \sin{\theta} \mathrm{d} \theta \int_0^{2\pi} \mathrm{d} \phi\ Z_{11}(\theta, \phi).
\end{equation}
Since we consider three additional wavelengths in the main part of the paper, we additionally checked the scattering phase function for them.
At $2\,\mathrm{\mu m}$, we also find a strong forward peak, but it is less strong than for $\lambda~=~0.8\,\mathrm{\mu m}$ by several orders of magnitude.
For the longer wavelengths of $12\,\mathrm{\mu m}$ and $21\,\mathrm{\mu m}$, there is only a very weak forward-scattering peak, which means that a consideration at these wavelengths does not require the chopping mechanism. However, the chopping mechanism also does not have a large effect on the results in these wavelengths. For simplicity, we therefore treat all simulations in the main part using the same opacity set-up with chopping of $2\degree$.

\begin{figure*}[ht!]
\sidecaption
  \includegraphics[width=12cm]{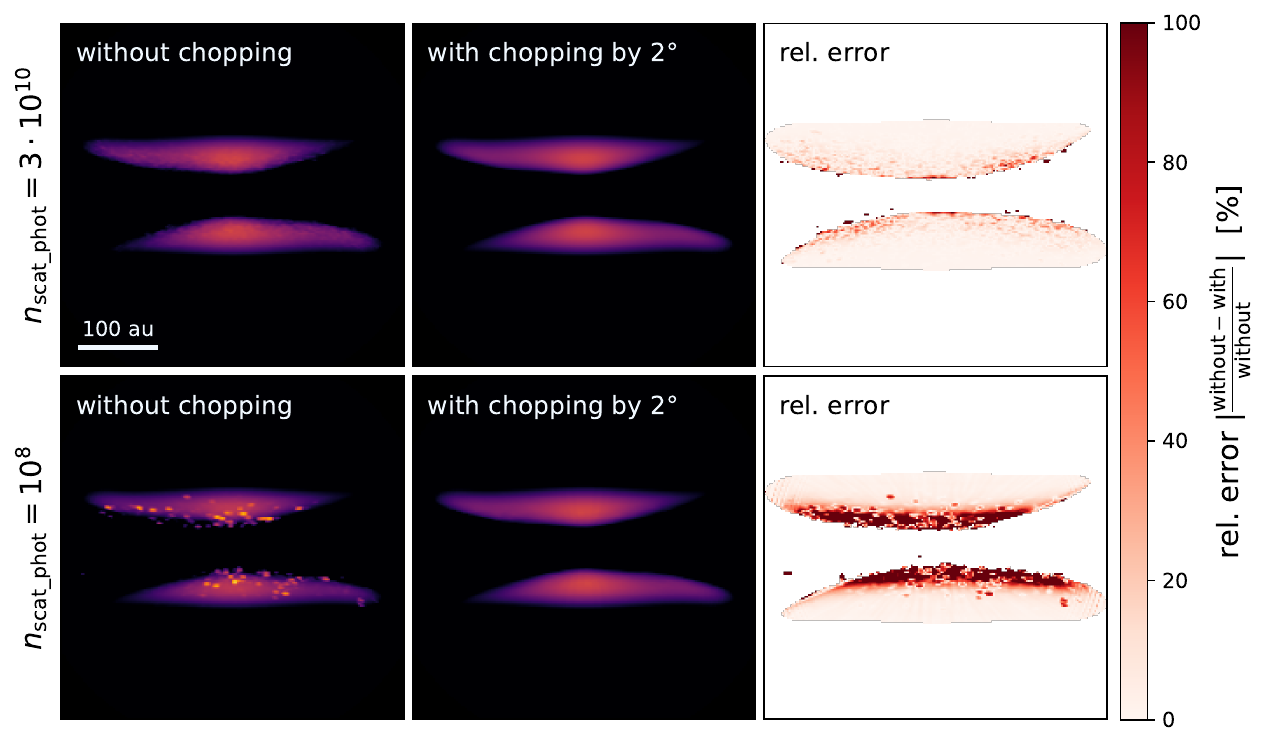}
  \caption{\label{fig:chopping} Comparison of the RADMC-3D images between cases with chopping by $2\degree$ and without chopping for $3 \cdot 10^{10}$ scattering photons (top row) and $10^8$ scattering photons (bottom row). These models are calculated at the wavelength $\lambda=0.8\,\mathrm{\mu m}$. The right column shows the relative error in percent. Here, we ignored intensity values of the images that are smaller than $10^{-17}$ in order to avoid artifacts.}
\end{figure*}

Interestingly, we noticed that for such extreme forward scattering peaks like in our case, the default internal angular resolution\footnote{The internal angular resolution $n_\mathrm{ang}$ can be set using the option \texttt{-s <n\_ang>} in \texttt{optool}.} of \texttt{optool} is not enough to accurately determine the scattering phase function.
We initially noticed a discrepancy between the phase functions with and without chopping using the default angular resolution $n_\mathrm{ang}=180$ of \texttt{optool}, see yellow and blue lines in Figure~\ref{fig:phase-function}.
We found that this difference in the scattering phase function leads to a relative difference of intensity in the images of about $30\%$.
We suspected the reason for this difference between the phase functions with and without chopping to be linked to the extreme forward scattering peak.
The tool \texttt{optool} normalizes the scattering matrix to produce the correct integral.
As the forward-scattering peak has a strong contribution to the integral, a slight inaccuracy within the peak could lead to strong inaccuracies for larger scattering angles.
Therefore, a higher internal angular resolution ensures an accurate calculation of the forward scattering peak, leading to a more accurate determination of the normalized values for high scattering angles.
On the other hand, if the extreme forward peak is cut off using the chopping mechanism, the normalization can be more accurate for high scattering angles in this case as well.
Indeed, using a higher internal angular resolution for the scattering phase function gets rid of this difference, as shown in Figure~\ref{fig:phase-function} in the green line, which almost perfectly aligns with the blue line in the regime of higher scattering angles.
Changing the internal angular resolution for the phase function with the chopping mechanism does indeed not significantly change the phase function (dotted line), as the normalization integral is sensitive to slight deviations when the peak is chopped.
We therefore conclude that it is not necessary to use the higher angular resolution for simulations when using a chopped phase function.

In order to ensure that the chopping does not change the correctness of the result, we ran comparison simulations using different amounts of scattering photons in a test case.
Because we performed this test in an early stage of this project, our test case has slightly different disk parameters.
We used a warped disk with a disk tilt of $i_\mathrm{max}=10\degree$ at a warp location of $r_\mathrm{warp}=20\,\mathrm{au}$.
The range of the disk and the coordinate system are the same as for the simulations in the main part of this work, but the aspect ratio of the disk is $h_0=0.05$ at the reference radius $R_0~=~5.2\,\mathrm{au}$ and the flaring index is $i_\mathrm{fl}=0.25$.
Additionally, we used a higher dust mass than in our main models, i.e. $M_\mathrm{dust}=10^{-4}M_*$, which leads to a different shape and separation width of the top and bottom nebula.

In total, we ran four simulations for this test: one set using $n_\mathrm{scat\_phot}=3\cdot 10^{10}$ scattering photons, and one set with $n_\mathrm{scat\_phot}=10^8$, each set both without chopping and with chopping of $2\degree$.
For the simulations without chopping, we used an internal angular resolution of the phase functions of $n_\mathrm{ang}=720$, while for the simulations with chopping, we kept the default resolution of $n_\mathrm{ang}=180$, since we were aiming at performing the final simulations using this default resolution.

\begin{figure} [ht!]
  \centerline{\includegraphics[width=0.45\textwidth]{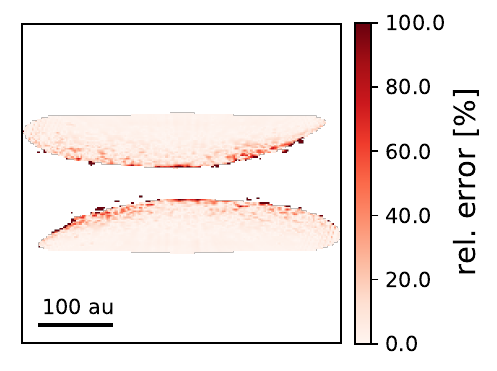}}
  \caption{\label{fig:chopping-relerror} Relative error between the model closest to reality ($n_\mathrm{phot\_scat}~=~3\cdot10^{10}$ without chopping, top left panel in Fig.~\ref{fig:chopping}) and the computationally least expensive model ($n_\mathrm{phot\_scat}=10^{8}$ with chopping by $2\degree$, bottom middle panel in Fig.~\ref{fig:chopping}).}
\end{figure}

Figure~\ref{fig:chopping} shows the image results for all four simulations, and in addition the relative error of the resulting intensities $ \delta = \left|\frac{I_\mathrm{without\ chopping}-I_\mathrm{with\ chopping}}{I_\mathrm{without\ chopping}}\right|$ for both sets.
In order to avoid artifacts in the residuals due to very small values in the images, we ignored the areas with intensity values lower than $10^{-17}$.
The bottom left panel of Figure~\ref{fig:chopping} shows the bright spots if too few scattering photon packages are used in the case without the chopping mechanism. The bright spots especially occur in the transition regimes between the optically thick and optically thin part of the disk.
They are unphysical artifacts due to the strong forward scattering peak at this wavelength.
As expected, there are two ways to avoid these bright spots. One way is to use more scattering photons (top left panel), which is physically closer to reality, but has the problem of long computation time.
The other way is to use the chopping mechanism (middle column), where we achieve a similarly good result at lower computational cost.

In Figure~\ref{fig:chopping-relerror}, we show the relative error between the model closest to reality and the trade-off model with low computational cost.
The median of the relative error is $2.7\%$, and we therefore conclude that it is safe to use a chopping of the phase function of $2\degree$ in order to save computation time for our final models.

\section{Unwarped disk} \label{sec:unwarped}

As a reference case for the models of warped disks, we investigate a planar, untilted disk in {near-/mid-infrared}.
As described in Section~\ref{sec:nummodels}, our disk contains dust of grain size $a=10\,\mathrm{\mu m}$.
Because this model is perfectly axisymmetric, we do not need to consider different viewing angles.

\begin{figure} [ht!]
  \centerline{\includegraphics[width=0.4\textwidth]{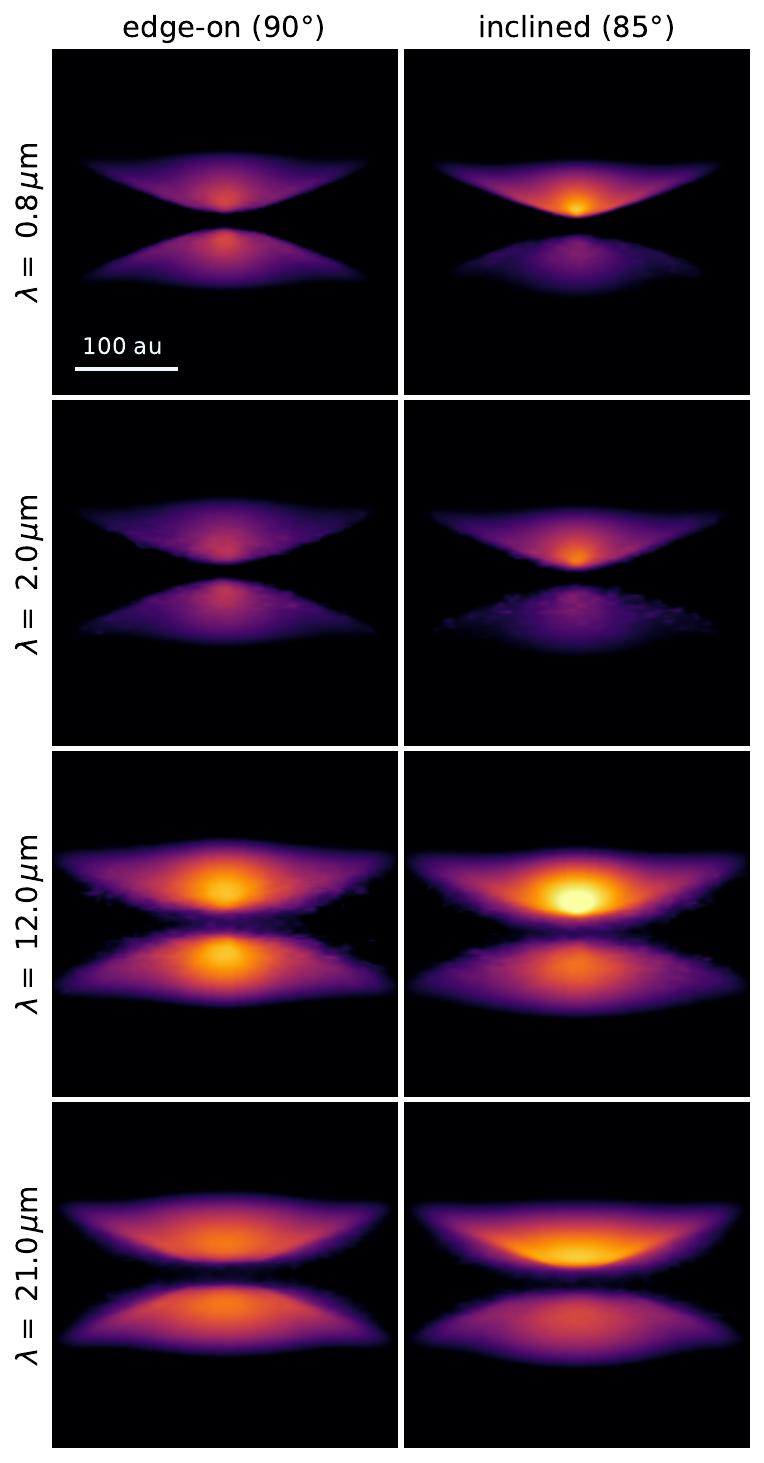}}
  \caption{\label{fig:unwarped} Synthetic {near-/mid-infrared radiative transfer} images {in log-scale} in different wavelengths of an unwarped disk. The left column shows the disk observed perfectly edge-on, the right column viewed at $85\degree$.}
\end{figure}

For an observer viewing the disk perfectly edge-on (inclination of $90\degree$), top and bottom nebulae are perfectly symmetric across multiple wavelengths, as shown in Figure~\ref{fig:unwarped}, top row.
Here, nebulae have the same shape and brightness.
Looking at the disk at a slight inclination of $85\degree$, the top nebula appears brighter in all wavelengths (see bottom row of Figure~\ref{fig:unwarped}).
This occurs mainly due to forward scattering, but backward scattered light from the far side of the disk contributes as well.

\section{{Testing smaller grain sizes}} \label{sec:grainsizes}

{In this section, we explore the lateral asymmetry and flux ratio in models with smaller dust grain sizes in the example of a \revnew{misalignment} of $i_\mathrm{max}=3^\circ$ and a warp location of $r_\mathrm{warp}=20\,\mathrm{au}$. In the main part, we focused on a single dust grain size of $a = 10\,\mu \mathrm{m}$. Here, we extend our investigations to $a = 1\,\mu \mathrm{m}$ and $0.5\,\mu \mathrm{m}$, as well as a dust size distribution for comparison. For the dust size distribution, we chose dust sizes from $a = 0.1\,\mu \mathrm{m}-10\,\mu \mathrm{m}$, distributed in a commonly assumed power-law distribution according to $n(a)\mathrm{d}a \propto a^{-3.5} \mathrm{d} a$. All other parameters remain the same as in the main part of this work.}

\begin{figure}[ht!]
    \centering
    \includegraphics[width=0.4\textwidth]{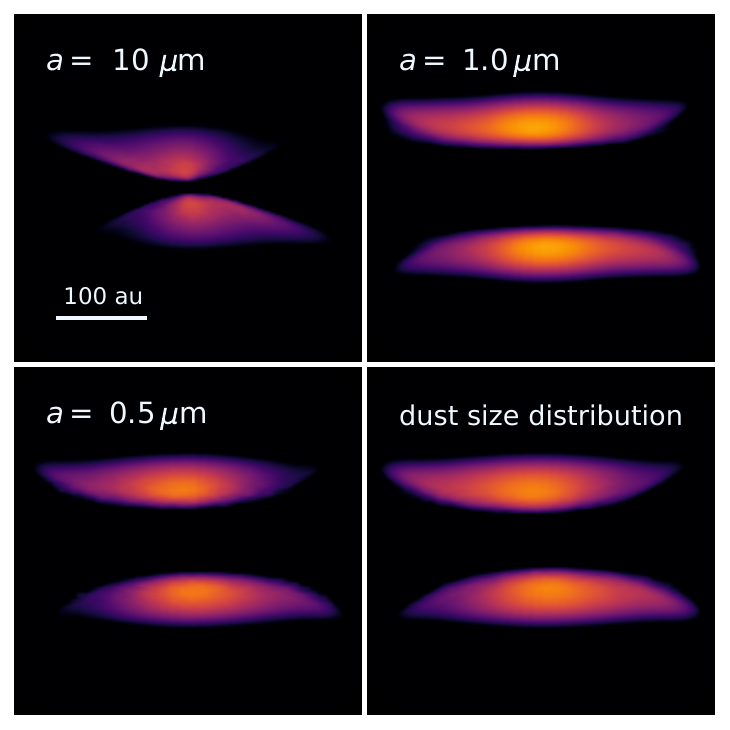}
    \caption{{Radiative transfer simulations of the model with a \revnew{misalignment} of $3^\circ$ at a location of $20\,\mathrm{au}$, viewed from $\varphi=0^\circ$, for different dust sizes. Images are shown for $\lambda=\,\mathrm{0.8}\,\mu\mathrm{m}$.}}
    \label{fig:dust-size-appearance}
\end{figure}

\begin{figure}[ht!]
    \centering
    \includegraphics[width=\linewidth]{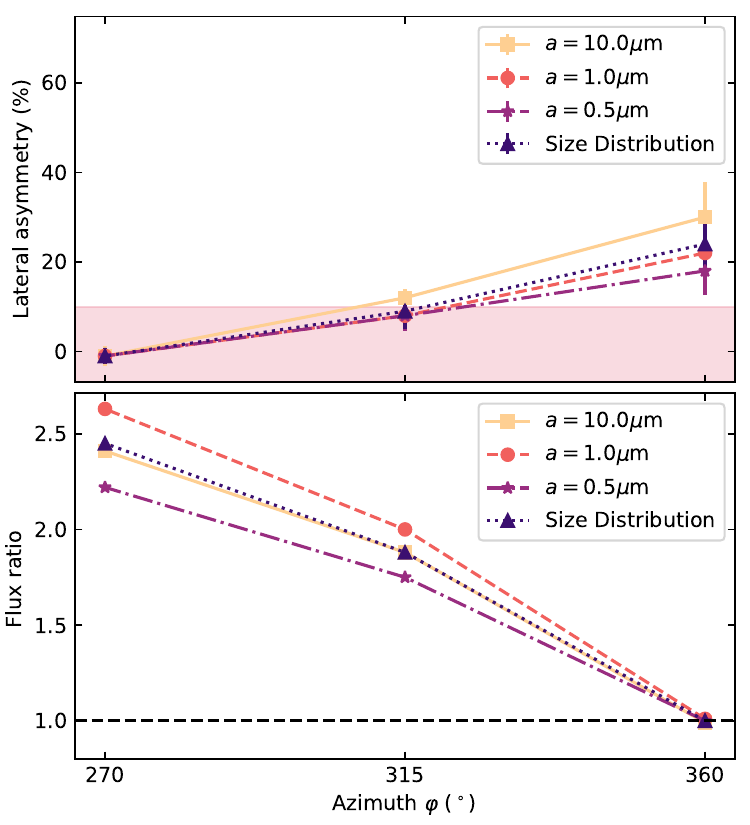}
    \caption{{Same as Figure~\ref{fig:spine_flux_strength}, but for a \revnew{misalignment} of $i_\mathrm{max}=3^\circ$ and different assumptions on the dust size in the models. All data points are for radiative transfer models at the wavelength $\lambda=0.8\,\mu\mathrm{m}$}.}
    \label{fig:dustsize}
\end{figure}

{We find that the appearance of the disk nebulae in the radiative transfer images differs between the models of different dust sizes, as shown in Figure~\ref{fig:dust-size-appearance}.
In especially, the shape of the nebulae is less rounded and the separation between the nebulae is wider for the smaller dust grain sizes.}
{This is because of the different opacities due to the different models of dust grain sizes.
However, when comparing the lateral asymmetry and flux ratio, we find the same trend as in our main models, as shown in Figure~\ref{fig:dustsize}.
This good agreement between the different assumptions on the dust
indicates that the results of this work in terms of lateral asymmetry and flux ratio can be applied in a more general way.} 

\begin{figure}[ht!]
    \centering
    \includegraphics[width=0.45\textwidth]{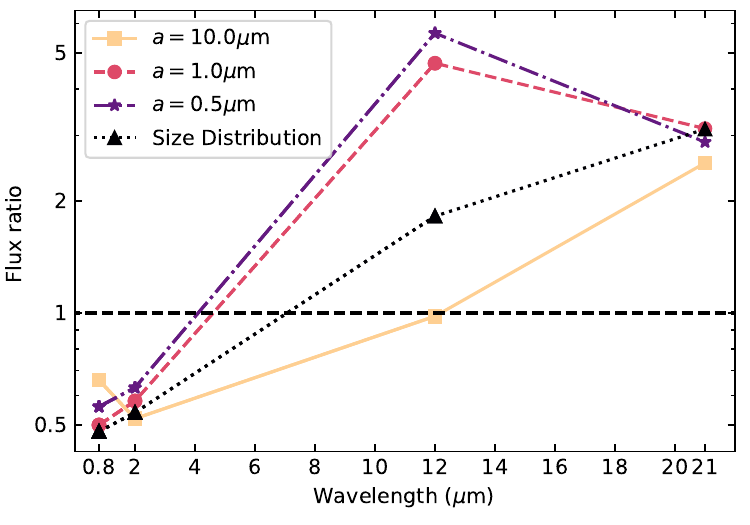}
    \caption{{Flux ratio as a function of wavelengths for models of warped disks with a warp tilt of $10^\circ$ and different dust properties. The models are simulated with $\varphi = 60^\circ$, and the $y$-axis is plotted in log-scale.}}
    \label{fig:wavelengthswap-differentDust}
\end{figure}

\begin{figure}[ht!]
    \centering
    \includegraphics[width=0.4\textwidth]{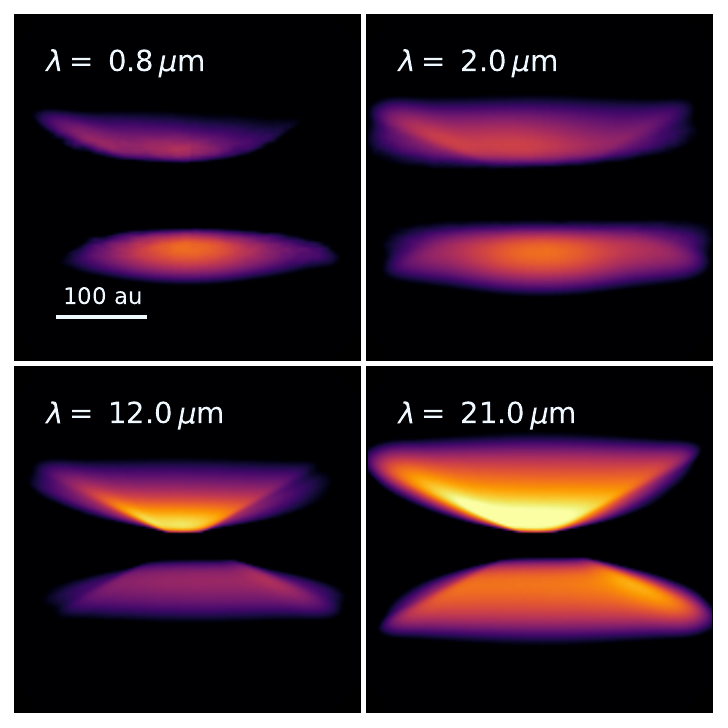}
    \caption{{Same as Figure~\ref{fig:tilt10-lambda}, but for the models with a dust grain size of $a = 1\,\mu \mathrm{m}$ instead of $10\,\mu \mathrm{m}$.}}
    \label{fig:wavelengthswap-1microndust}
\end{figure}

{We further checked for the different dust size models, if the swap in flux ratio with wavelength (see Sect~\ref{sec:multiple-wavelengths}) can be reproduced for smaller grains. Figure~\ref{fig:wavelengthswap-differentDust} shows that all three dust size models can reproduce the swap in the brightest nebula with wavelength. In particular, for all alternative dust models explored in this section, the flux ratio swap occurs at shorter wavelength than in the case of our fiducial model with $a=10\mu$m, which is noticeable at 12$\mu$m instead of 21$\mu$m. We show in Figure~\ref{fig:wavelengthswap-1microndust} the images at the different wavelengths for a dust size of $a = 1\,\mu \mathrm{m}$.}

{We note that for our chosen fiducial dust size of $a~=~10\,\mu\mathrm{m}$, the silicate feature in the absorption opacity at roughly $\lambda=10\,\mu\mathrm{m}$ is weak. However, the other models shown here have a stronger silicate feature because of the smaller grains.
Checking the appearance of the images for wavelengths around $\lambda~=~9~-~12\,\mu\mathrm{m}$, we found that the silicate feature does not have a strong effect on the lateral asymmetry.}

\end{appendix}

\end{document}